\documentclass[sigconf,nonacm]{acmart} 

\AtBeginDocument{%
  \providecommand\BibTeX{{%
    \normalfont B\kern-0.5em{\scshape i\kern-0.25em b}\kern-0.8em\TeX}}}

\setcopyright{rightsretained}
\usepackage{soul}
\usepackage{listings}

\begin{document}

\title{NL2KQL: From Natural Language to Kusto Query}

\author{Xinye Tang}
\authornote{Both authors contributed equally to this research.}
\email{xinye.tang@microsoft.com}
\author{Amir H. Abdi}
\authornotemark[1]
\email{amirabdi@microsoft.com}
\affiliation{%
  \institution{Microsoft} 
  \country{USA, Canada}
}

\author{Jeremias Eichelbaum}
\email{jeichelbaum@microsoft.com}
\affiliation{%
  \institution{Microsoft}
  \country{USA}
  }

\author{Mahan Das}
\email{mahandas@microsoft.com}
\affiliation{%
 \institution{Microsoft}
 \country{USA}
}

\author{Alex Klein}
\email{alexklein@microsoft.com}
\affiliation{%
  \institution{Microsoft}
  \country{USA}
}

\author{Nihal Irmak Pakis}
\email{npakis@microsoft.com}
\affiliation{%
  \institution{Microsoft}
  \country{Canada}
  }

\author{William Blum}
\email{william.blum@microsoft.com}
\affiliation{%
  \institution{Microsoft}
  \country{USA}
  }

\author{Daniel L Mace}
\email{daniel.mace@microsoft.com}
\affiliation{%
  \institution{Microsoft}
  \country{USA}
  }

\author{Tanvi Raja}
\email{tanvi.raja@microsoft.com}
\affiliation{%
  \institution{Microsoft}
  \country{USA}
  }

\author{Namrata Padmanabhan}
\email{npadmanabhan@microsoft.com}
\affiliation{%
  \institution{Microsoft}
  \country{USA}
  }

\author{Ye Xing}
\email{yexing@microsoft.com}
\affiliation{%
  \institution{Microsoft}
  \country{USA}
  }
\renewcommand{\shortauthors}{Tang and Abdi et al.}

\begin{abstract}

Data is growing rapidly in volume and complexity. Proficiency in database query languages is pivotal for crafting effective queries. As coding assistants become more prevalent, there is significant opportunity to enhance database query languages.
The Kusto Query Language (KQL) is a widely used query language for large semi-structured data such as logs, telemetries, and time-series for big data analytics platforms.
This paper introduces NL2KQL an innovative framework that uses  large language models (LLMs) to convert natural language queries (NLQs) to KQL queries.
The proposed NL2KQL framework includes several key components:
the \textit{Schema Refiner} which narrows down the schema to its most pertinent elements;
the \textit{Few-shot Selector} which dynamically selects relevant examples from a few-shot dataset;
and
the \textit{Query Refiner} which repairs syntactic and semantic errors in KQL queries.
Additionally, this study outlines a method for generating large datasets of synthetic NLQ-KQL pairs which are valid within a specific database contexts.
To validate NL2KQL's performance, we utilize an array of online (based on query execution) and offline (based on query parsing) metrics.
Through ablation studies, the significance of each framework component is examined, and the datasets used for benchmarking are made publicly available. This work is the first of its kind and is compared with available baselines to demonstrate its effectiveness.

\end{abstract}

\ccsdesc[300]{Information systems~Language models}
\ccsdesc[500]{Computing methodologies~Natural language generation}
\ccsdesc[300]{Human-centered computing~Natural language interfaces}

\keywords{generative model, language model, kusto query language, KQL, code generation, automated data analytics}





\maketitle

\pagestyle{empty}

\section{Introduction}



Computer scientists have long been interested in the longstanding challenge of program synthesis and translating intent, expressed in natural language, into executable code, even before the advent of modern techniques such as machine learning, deep learning, and transformers~\cite{waldinger69,simon63,summers77}.
With the availability of large code datasets, Large Language Models (LLM) superseded classic sequence generation models and the likes of CodeBERT~\cite{Feng2020CodeBERTAP}, PyMT5~\cite{Clement2020PyMT5MT} and Codex~\cite{codex} became more prevalent.  These models led to advent of coding assistants which are gaining popularity each day. Research suggests that coding assistants enhance the productivity and learning speed of their users~\cite{productivity22}.

Query languages are feature-rich and powerful but often present a learning curve that not all users can easily master.
Consequently, there is an ever increasing research on Natural Language  (NL) Interface for Databases (NLIDB).
With Structured Query Language (SQL) being the most widely used query language,  many NLQ-to-SQL solutions have been proposed over the years ranging from parsing-based approaches, to more recent Neural Machine Translation (NMT) techniques~\cite{sqlsurvey}. 
The two widely used NLQ-SQL datasets, WikiSQL~\cite{wikisql} and Spider~\cite{spider}, 
and the recently released BIRD benchmark~\cite{bird}, have led to nearly a thousand NLQ-to-SQL solutions.

Unlike SQL, which deals with structured data hence the name, Kusto Query Language (KQL) is designed for large semi-structured data such as logs, telemetry data, and time-series, which are commonly found in big data analytics platforms. 
These characteristics are blessings for KQL as a powerful query language and a curse in converting NL to KQL.

In this study, we introduce an innovative approach for transforming Natural Language Queries (NLQ) into Kusto Query Language (KQL) queries, denoted as NL2KQL. 
This solution capitalizes on the capabilities of LLMs.
The generated KQLs are semantically valid for the target Kusto database using database-specific synthetic few-shots and a Semantic Data Catalog. 
The principal contributions of this research are multifold.
Firstly, we present the first end-to-end automated NLQ-to-KQL framework, incorporating novel methods for database-specific schema refinement and on-the-fly query refinement.
Secondly, we introduce a method to synthesize large scale NLQ-KQL pairs with high variance tuned for a Kusto database.
Thirdly, we propose a set of offline and online metrics to assess the effectiveness of our approach, some of which hold the potential for application in areas beyond KQL generation. 
Fourthly, we release the first benchmark for KQL generation evaluation, along with their Semantic Data Catalog, comprising 400 NLQ-KQL pairs compiled by KQL experts, distributed across two Kusto databases and segmented into easy and hard categories.
Lastly, an exhaustive ablation study is conducted to delineate each component's individual contributions to the overall performance.



\begin{figure*}
    \includegraphics[width=1\linewidth]{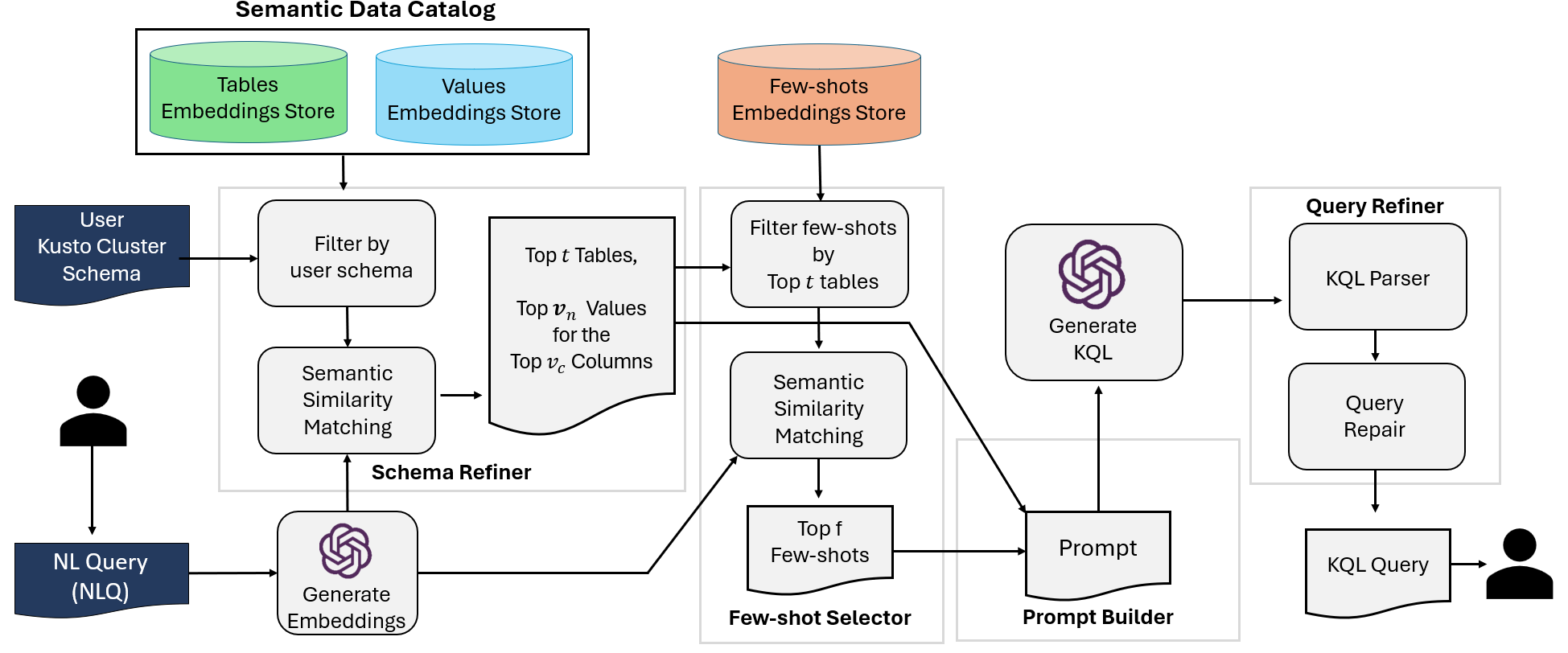}
    \caption{Overview of NL2KQL inference pipeline.}
    \label{fig:enter-label}
    \label{fig:teaser}
\end{figure*}

\section{Related Works}

\subsection{Retrieval-enhanced Generation}

Historical and recent studies have demonstrated that retrieval mechanisms to enhance performance of auto-regressive models, including question answering \cite{voorhees-tice-2000-trec, chen-etal-2017-reading,kwiatkowski-etal-2019-natural}, fact-checking \cite{thorne2018fever}, dialogue systems \cite{dinan2018wizard}, and citation recommendation \cite{logan2021cutting}. Initially, retrieval was predominantly conducted through term-matching techniques like TF-IDF and BM25 \cite{brown2020language}.
The advent of neural networks ushered the era of dense retrievers, adopting dual-encoder architectures~\cite{yih-etal-2011-learning,inproceedings,10.1145/2567948.2577348}.
Notable advancements include DPR, aimed at distinguishing relevant passages among non-relevant ones, and its enhancements like ANCE which refine the process of mining hard negatives~\cite{karpukhin2020dense,xiong2020approximate}. 
Yates et al. have composed a comprehensive overview of dense retrieval techniques~\cite{10.1145/3437963.3441667}.

After retrieval, the pertinent documents undergo processing to generate the final output. In open-domain question answering, approaches vary from extracting text spans from documents~\cite{chen-etal-2017-reading,clark-gardner-2018-simple,wang2019multipassage,karpukhin2020dense} 
to leveraging seq2seq models that generate answers in free-form text based on the retrieved documents~\cite{karpukhin2020dense,izacard2022distilling}. 
Notably, such architectures have also been instrumental in reducing hallucination in dialogue systems~\cite{shuster2021retrieval}.

\subsection{NLQ-to-SQL}
For a long while, rule-based methodologies and   custom heuristics dominated the NLQ-to-SQL domain~\cite{Christopher_2019, Fei10.14778/2735461.2735468, Diptikalyan10.14778/2994509.2994536, Jaydeep2020ATHENANL}. These methods typically employ established NLP techniques to deconstruct a question into a parsed tree. Based on the tree's structure, specific rules are formulated to transform this tree into an SQL Abstract Syntax Tree (AST), which is subsequently converted into the executable SQL query. 
Nonetheless, adapting these methods to new database domains often demands manual effort.

Recently,  sketch-based learning techniques have introduced more generalizable solutions where a SQL sketch or template with placeholders is employed.
These methods bypasses the need for generating SQL syntax, focusing instead on  extracting and inserting relevant information into the predefined slots~\cite{Xiaojun, Tao}. 

In contrast, sequence-to-sequence models view SQL queries as a series of tokens, applying deep learning techniques. The primary challenge here is ensuring the syntactical correctness of the output SQL sequence. Notably, PICARD~\cite{Torsten} represents an advanced solution in this domain, employing a syntax-validation mechanism during the beam search process 
ValueNet discusses use of values along with schema information to improve the performance~\cite{brunner2021valuenet}.
Katsogiannis et al. have composed a thorough  review of NLQ-to-SQL research~\citep{sqlsurvey}.

\section{Preliminary}

\subsection{Kusto Query Language (KQL)}
Kusto Query Language (KQL) is a powerful query language designed to analyze big data stored in Azure Data Explorer, a scalable data exploration service from Microsoft Azure\footnote{https://learn.microsoft.com/en-us/azure/data-explorer/kusto/query/}. KQL enables users to extract, manipulate, and analyze large datasets, making it more than just a query language but a tool for insights. 
Unlike traditional SQL, KQL is optimized for unstructured and semi-structured data like logs, metrics, and events, which are common in modern applications and systems.

Similar to SQL databases, a kusto database is defined by its schema which contains the names of tables and their respective column names and column types.


Core constructs of KQL include Identifiers (names of columns, tables, and variables), Literals (fixed values), and Expressions, where each Expression constitutes of an operator 
(arithmetic operators, comparison operators, logical operators, string operators, and aggregation operators) and its operands.

Despite its simplicity, KQL is a powerful data analytics tool with flexible syntax and various forms and constructs beyond the pattern. Further details on these query forms are discussed in Appendix~\ref{apx:kql-forms}. 


\subsection{NLQ-to-KQL Problem Statement}

The NLQ-to-KQL problem is formalized as:
Given a Natural Language Query (NLQ) and a Kusto database with a known schema~($s$), generate a KQL query~($\hat{q}$) which is valid with respect to $s$ and, if executed, fetches results that match the user's intent.

\subsection{NLQ-to-KQL Challenges}
\label{sec:challenges}

Like all other programming and query languages, KQL has a strict syntax with less expressivity than natural language (NL) which leads to a non-bijective translation.
The inherent ambiguity of NL such as 
lexical ambiguity (word with multiple meanings) and
syntactic-syntactic ambiguity (sentence with multiple interpretations), 
make the translation even more so challenging~\cite{sqlsurvey}.
An NLQ can be paraphrased in various ways, may lack key terms, or contain syntactical or grammatical mistakes.
In a chat user interface with conversation history, user's follow-up questions add yet another layer of complexity.

On the other hand KQL is not forgiving when it comes to syntactic or semantic errors. The generative model has to have knowledge of the Kusto database schema which may not be well-structured. KQL is designed for logs, time-series and telemetry, focusing on data exploration and analytics. 
KQL supports joining tables based on matching values, however, unlike SQL, 
it has no notion of primary and foreign keys and does not enforce referential integrity or uniqueness constraints.
Complicating matters, commonly, KQL queries parse JSON columns to extract keys and the extracted JSON key, in turn, can be used to join tables~(see Appendix~\ref{sec:parse-json-join}).
Furthermore, telemetry data is unnormalized, inherently redundant, with columns duplicated across tables, all of which are confusing to an auto-regressive generative language model.

\section{Proposed Method}

The inherent complexities and ambiguity of natural language, plus the specificity and flexibility of KQL, are the main drivers of the design behind the proposed framework to convert Natural Langugae Query (NLQ) to Kusto Query Language (KQL), namely the \textit{NL2KQL}.
This end-to-end system is composed of a sequence of components as followed:
\textit{Semantic Data Catalog}, 
\textit{Schema Refiner}, 
\textit{Few-shot Database}, 
\textit{Few-shot Selector}, 
\textit{Prompt Builder}, and 
\textit{Query Refiner}~(Figure~\ref{fig:teaser}). 
The subsequent sections will delve into the details of each component.

\subsection{Semantic Data Catalog}
\label{sec:data-catalog}

\begin{figure}
    \centering
    \includegraphics[width=1\linewidth]{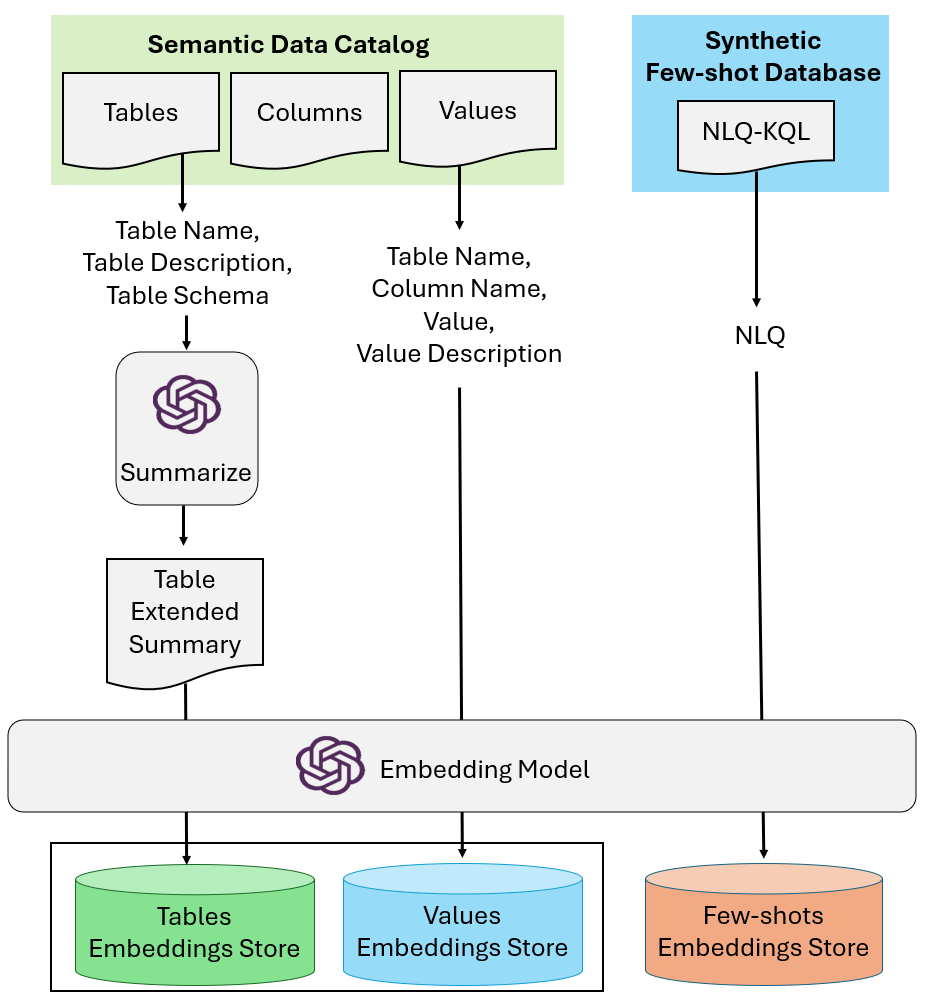}
    \caption{Overview of embedding stores}
    \label{fig:embedding}
\end{figure}

A \textit{Semantic Data Catalog} plays a pivotal role in encapsulating information about the structure, semantics, and contextual attributes of a database. 
The \textit{Schema Refiner} and \textit{Query Refiner} both leverage \textit{Semantic Data Catalog} and the embeddings of its elements to make decisions for schema filtering and query repair~(Figure~\ref{fig:embedding}).



Semantic Data Catalog provides annotations for tables, columns, type constraints, and, where applicable, enumerated values. The schema of the Semantic Data Catalog is as followed:
\begin{lstlisting} [basicstyle=\small, frame=single, breaklines=true, columns=fullflexible]
- Table:
  - Name: string
  - Description: string
  - Columns:
      - Column:
          - Name: string
          - Type: string
          - Description: string
          - Format: Enum (optional)
          - Values: (optional)
              - Value:
                  - Value: string
                  - Description: string (optional)
\end{lstlisting}





The Semantic Data Catalog
is created via an automated process.
For a given Kusto database, the schema structure
are retrieved, either directly from the database
or a third-party API, and 
schema descriptions are extracted from 
Azure Monitor Data Reference~\footnote{https://learn.microsoft.com/en-us/azure/azure-monitor/reference/} 
Generic data retrieval queries are executed on the Kusto database to obtain a sample of data for each table, allowing for the extraction of unique values for the low-cardinality Enum columns. 
Once elements are structured, an \textit{extended summary} is created for each as followed:\\
\noindent\textbf{Table Element:} A generative LLM is tasked to use table name, table description, and table schema, including columns, to write an extended summary of the table covering scenarios where this table would be useful.\\
\noindent\textbf{Value Element:} Concatenation of table name, column name, value, and the value description.

The embeddings of the extended summaries of all elements are calculated via a similarity embedding model (Figure~\ref{fig:embedding}) and stored in a vector database for use at inference time. 





\subsection{Schema Refiner}

The model requires a representation of the schema to generate a semantically valid KQL.
Including Schema in the context of the model further helps ground LLMs and reduce the known risk of hallucination.
However, incorporating the entire database schema into the prompt poses challenges.
Several key considerations underscore the limitations of including the entire schema:

\subsubsection*{Finite Model Context Window:}
A common challenge of LLMs is their limited context window size which can render insufficient to hold the entire schema.
Therefore, a generalizable NLQ-to-KQL system should take smarter decisions for schema representation.



\subsubsection*{Permission Constraints:}
Not all customers have access to all tables in a given schema. Generating queries for tables inaccessible by the customer is semantically incorrect. 

In response to these challenges, and similar to the Schema Linking practices of NL-to-SQL~\cite{text2sqlbenchmark}, the \textit{Schema Refiner} selects most relevant tables, columns, and potential values, and  incorporates them 
in model's context. 
The refiner relies on similarity of NLQ embeddings with elements of the Data Catalog. 


During inference, the schema is initially filtered based on the user's permissions.
Then, a maximum of $t$ similar tables are identified by calculating the cosine similarity between their embeddings and the NLQ embedding, where $t$ is a predefined hyper-parameter.
All the columns in the selected tables are included in the schema.
As discussed in Section~\ref{sec:data-catalog}, for some low-cardinality columns, the list of acceptable values are included in the prompt. However, to not pollute the context with irrelevant tokens, only the top $v_{\text{n}}$ values of a column whose embedding have close similarity with the NLQ embedding would be included in the schema. 

\subsection{Synthetic Few-shot Database}

 \begin{figure}
     \centering
     \includegraphics[width=1\linewidth]{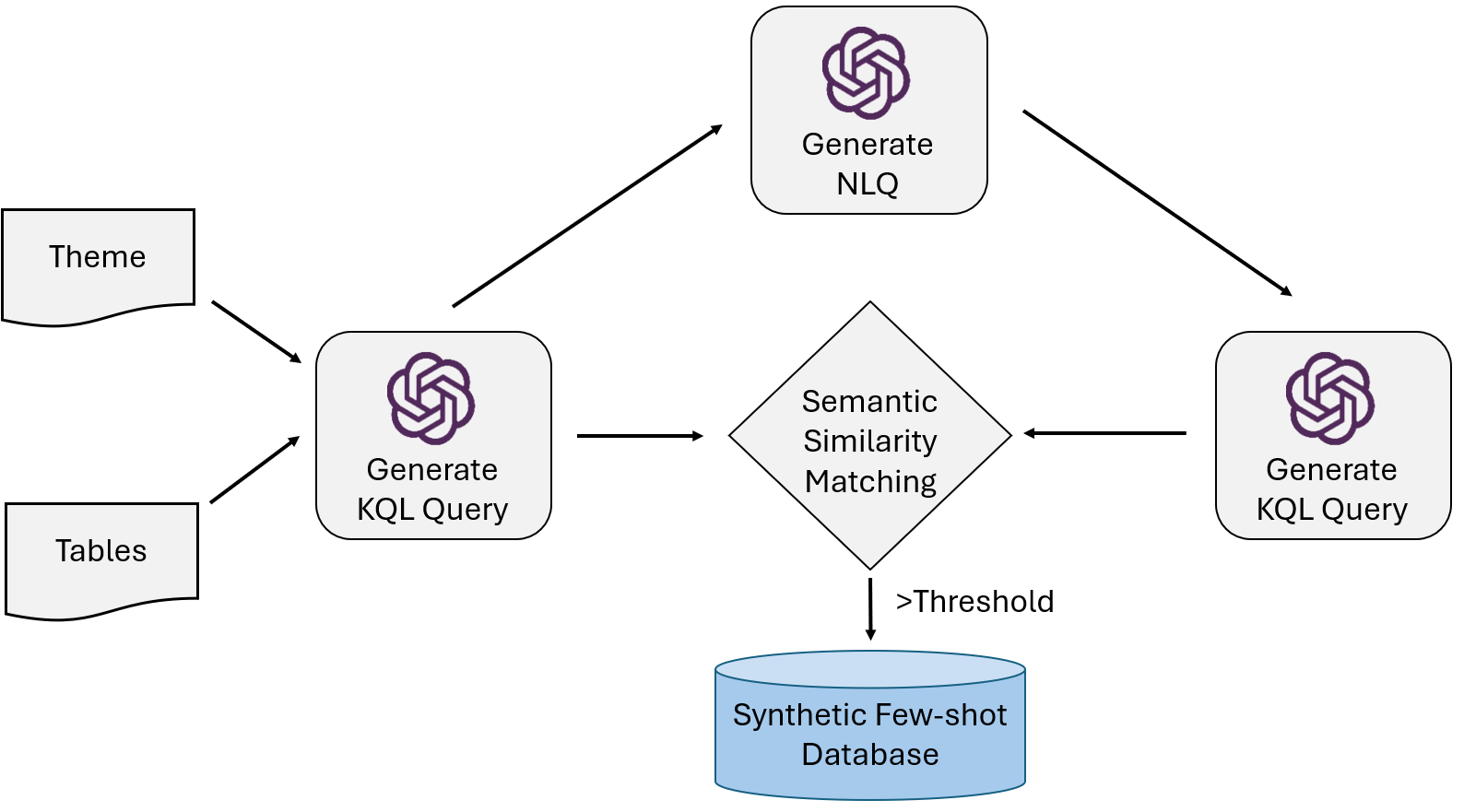}
     \caption{Synthetic few-shot generation and round-trip validation process.}
     \label{fig:roundtrip}
 \end{figure}

The NL2KQL solution uses a synthetic few-shot database (FSDB) to guide the LLM. 
The FSDB is synthesized for a target Kusto database and contains KQLs which are valid with respect to the database's schema.
The generated KQLs go through 3 rounds of validation to ensure sanity, syntactic and semantic correctness, and to filter out any ambiguous NLQs.

The synthetic KQL generation starts with random table sampling from the database schema. 
For 70\% of queries only one table is sampled.
For the remaining 30\%, a second table is chosen for a join, with half of these selected based on similar column types.
The final set of tables (i.e., one table without join, and two tables with join) are passed to the next step (Figure~\ref{fig:roundtrip}).

Similarly, a theme is randomly sampled from the following five themes: 
1) Explore: Look for signs or hints of a security attack;
2) Expansion: Searches for additional contextual understanding for the scenario;
3) Detect: Look for events related to a security attack;
4) Remediate: Identify all evens for a given entity or asset;
5) Report: Provide summary statistics that helps with writing a report.
The inclusion of random themes  further promotes the variability of the final KQLs.

An LLM generate a KQL query given the table set (schema) and the theme (Figure~\ref{fig:roundtrip}). 
The prompt contains instructions on Kusto syntax with query design options. 

Once the KQL query is generated, the same LLM is tasked to explain the KQL in imperative  natural language (NLQ). 
This process is repeated with the NLQ included in the prompt to generate a secondary KQL query. Both primary and secondary KQLs are validated with KQL parser for syntactic and semantic correctness and disregarded if they contain errors.

The initial and secondary KQLs are compared using the Jaccard similarity of their tokens and rejected if their similarity falls below the 0.7 point. 
Accepted primary KQLs and their corresponding NLQs are added to the FSDB, ensuring that each NLQ accurately represents its KQL query without ambiguity.










\subsection{Few-shot Selector}

Large language models are proven to be few-shot learners. 
The best few-shots are those which introduce nuanced hints to the model 
and avoid  misleading cues. 
Therefore, \textit{Few-shot Selector} dynamically selects few-shots with respect to the NLQ and the user's context while considering the following:


\subsubsection*{Number of Few-shots:}
Previous research has demonstrated that Few-shots tend to outperform zero-shot scenarios. It has been established in literature  that employing a tiny number of few-shots is sufficient and additional few-shots do not yield further performance improvements~\cite{fewshot3, gpt2020}.

\subsubsection*{Schema-relevance of Few-shots:}
Syntactic correctness is a necessary but insufficient condition for the few-shots. Ensuring that few-shots only contain tables and columns within the accessible schema of the customer is vital to lead the LLM into generating semantically correct KQLs.



Considering the above, a few-shot database (FSDB) is created where each sample is an NLQ-KQL pair. The embeddings of the NLQ part of all few-shots are calculated offline and stored in a vector database for use at inference time~(Figure \ref{fig:embedding}).
During inference, given a user NLQ, the FSDB is primarily filtered to only include schema-relevant few-shots. 
The  embeddings of the filtered few-shots are then compared with the embeddings of user's NLQ and the top $f$ few-shots are selected based on their cosine similarity.



\subsection{Prompt Builder}
Crafting an effective prompt for LLM depends on careful integration of its components, each contributing to the contextual richness autoregerssive generation process. The NL2KQL's prompt for the pre-trained LLM has the following elements:

\subsubsection*{Instructions:}
The instructions section sets the stage by providing essential scenario information. It outlines LLM's role, the nature of the user's request, and the expected output. This section serves as a guiding framework for the LLM and instructs the model to follow a Chain-of-Thoughts reasoning~\cite{CoT}.

\subsubsection*{Schema}
The schema section includes the narrowed-down schema extracted by the \textit{Schema Refiner}. It consists of tables, columns, and potential values.


\subsubsection*{Kusto Syntax:}
This section presents selected, useful Kusto syntax elements to influence generation of syntactically correct and semantically meaningful queries. 
It encompasses Scalar Functions, Aggregation Functions, Window Functions, Tabular Operators, and Scalar Operators. 

\subsubsection*{Kusto Best Practices:}
This section provides general guidelines and recommendations for optimized, efficient, and maintainable queries.
It covers best practices for operators, aggregations, joining tables, and includes additional tips and tricks. 

\subsubsection*{Few-shots:}
It includes the top $f$ few-shots selected by the \textit{Few-shot Selector}.

Template of the meta prompt is available in Appendix \ref{apx:prompt-template}.

\subsection{Query Refiner}
The \textit{Query Refiner} is a post-processor of the generated KQL query 
to check syntactic and semantic  correctness and, if possible, repair them. 
It employs the official KQL parser library
which detects syntactic and semantic errors given the schema~\footnote{https://github.com/microsoft/Kusto-Query-Language}.
With its recursive approach, the \textit{Query Refiner} handles undefined variables, fixes join statements, adds missing operators, among other foreseen errors.

To detect undefined identifiers, \textit{Query Refiner} uses embeddings of columns' and tables' descriptions, with a strict cosine similarity threshold of 0.9, to find substitute identifiers.


Besides accounting for undefined identifiers, \textit{Query Refiner} applies a few more rule-based fixes such as:
1) add the missing \textit{summarize} operator when aggregate functions are used improperly, 
2) add properly paired parentheses for the \textit{between} operator, and
3) add missing \textit{extend} operator.

\section{EXPERIMENTS}

All the experiments were conducted with two pre-trained LLMs: OpenAI's gpt-4-32k-0613, with a context size of 32k and a Chat Completion interface, was used for generative tasks and OpenAI's similarity embedding model, text-embedding-ada-002, was used to generate embeddings. 

Moreover, to create a more challenging environment with a reasonably large schema size, it was assumed that the user has access to all the tables in the Kusto database's schema.
During Schema Refinement, maximum number of tables, $t$, was set to 9. 
Maximum number of values per each Enum column, $v_{\text{n}}$, was set 5.
The number of few-shots, $f$, was set to 2. The FSDB includes 16 few-shots for Defender and 19 for Sentinel.

The only source of randomness in the experiments stemmed from the LLM completions. Despite conducting completions at zero temperature to minimize variability, determinism cannot be guaranteed due to the inherent non-determinism of asynchronous floating-point operations on GPUs. To address potential fluctuations, experiments were repeated 3 times and outcomes were averaged for consistency.

\subsection{Evaluation Benchmark}

The proposed NL2KQL solution was evaluated on two Kusto databases with different schemas. 

\textbf{Sentinel}: The first schema originated from a Kusto database of Microsoft Sentinel, a Security Information and Event Management (SIEM) solution  that delivers intelligent security analytics and threat intelligence across the enterprise.
The Sentinel database has 23 tables.
A team of security operations center analysts drafted 197 NLQ-KQL pairs targeted at the Sentinel database which are entirely used as the evaluation set.

\textbf{Defender}: The second schema is from Microsoft Defender XDR, an Extended Detection and Response (XDR) solution that coordinates detection, prevention, investigation, and response across endpoints, identities, email, and applications pre- and post-breach.
The Defender database has 29 tables.
Following the same process as the other database, 230 NLQ-KQL pairs were drafted as the Defender evaluation set.

In both databases, the security analysts were tasked to write NLQ-KQL pairs that represent their day-to-day queries without including any Personally identifiable information (PII).
The analysts had zero exposure to the Few-Shot Databases to mitigate any unintended bias and information leakage.


The two evaluation datasets for Defender and Sentinel, along with their respective Semantic Data Catalogs, are published and publicly accessible at https://github.com/xinyeah/NL2KQL. 
A summary of the schema, including list of tables and columns are available in Appendix~\ref{apx:schema}.

\subsection{Evaluation Metrics}

In each NLQ-KQL evaluation sample, the KQL is considered the ground-truth target, $q$, for the input NLQ. 
During evaluation, the predicted (generated) KQL query, $\hat{q}$ is evaluated against $q$.
Given the intricacies of KQL, a portfolio of metrics are designed which are grouped under two families: Offline metrics and Online metrics.

\subsubsection{Offline Metrics}
These metrics focus on the syntactic correctness of $\hat{q}$ as well as the syntactic and  semantic similarities of $\hat{q}$ and $q$.
To capture and compare various characteristics of $q$ and $\hat{q}$ the following offline metrics are designed.

\noindent\textbf{Syntax Score:} Evaluates syntactic correctness of $\hat{q}$.
\[
\text{Syntax}(\hat{q}) = 
\begin{cases} 
1 & \text{if } \hat{q} \text{ is syntactically correct,} \\
0 & \text{otherwise.}
\end{cases}
\]
    
\noindent\textbf{Semantic Score:} Evaluates semantic correctness of $\hat{q}$ within the context of the schema, $s$.
\[
\text{Semantic}(\hat{q}, s) = 
\begin{cases} 
1 & \text{if } \hat{q} \text{ is semantically correct,} \\
0 & \text{otherwise.}
\end{cases}
\]

\noindent\textbf{Table Score:} Evaluates the proportion of tables referenced in the \(\hat{q}\) query ($T(\hat{q})$) which are also references in \(q\), but only if \(T(q)\) is a subset of \(T(\hat{q})\); otherwise, it's zero.
\[
\text{Table}(q, \hat{q}) = 
\begin{cases} 
\frac{|T(q) \cap T(\hat{q})|}{|T(\hat{q})|} & \text{if } T(q) \subseteq T(\hat{q}), \\
0 & \text{otherwise.}
\end{cases}
\]

\noindent\textbf{Filter Column Score:} Evaluates Jaccard similarity of the set of columns referenced in filters ($F_{\text{col}}(.)$) of \(\hat{q}\) and \(q\), where Jaccard of two sets is defined as $Jaccard(a,b)=\frac{|a \cap b|}{|a \cup b|}$.
\[
\text{Filter}_{\text{col}}(q, \hat{q}) = 
Jaccard(F_{\text{col}}(q), F_{\text{col}}(\hat{q})) .
\]

\noindent\textbf{Filter Literal Score:} Evaluates Jaccard similarity of the set of literals used in filters 
of \(\hat{q}\) and \(q\).






\subsubsection{Online Metrics}

The inherent data redundancy of logs and telemetries in Kusto databases and KQL's flexibility in expressing the same query in different ways erodes uniqueness of ground-truth KQL that corresponds to a given NLQ. Online metrics, which measure the similarities of the query results, are intended to address the limitations of offline metrics.
The online metrics are as followed.

\noindent\textbf{Row Execution Score:} Evaluates Jaccard similarity of records (rows) in the results retrieved by $q$ and $\hat{q}$.
\[
\text{Row}_{\text{exec}}(q, \hat{q}) = 
Jaccard(R_r(q), R_r(\hat{q})) .
\]

\noindent\textbf{Column Execution Score:} Evaluates the proportion of columns in the results returned by \(\hat{q}\) query which are also returned by \(q\).
\[
\text{$\text{Column}_{\text{exec}}$}(q, \hat{q}) = \frac{|C_r(q) \cap C_r(\hat{q})|}{|C_r(\hat{q})|} .
\]

\noindent\textbf{Execution Score:} Average of column and row execution scores.
\[
\text{Avg}_{\text{exec}}(q, \hat{q}) = \frac{\text{Row}_{\text{exec}}(q, \hat{q}) + \text{Column}_{\text{exec}}(q, \hat{q})}{2} .
\]


\section{Results and Analysis}

The proposed method is compared with the available baselines across the 8 mentioned metrics.
A thorough ablation study evaluate the performance of each component and their alternatives.
The results of offline and online metrics are summarized in Table~\ref{tab:main_results} and Table~\ref{tab:main_results_online}, respectively.
Across all metrics, we have found the $\text{Avg}_{\text{exec}}$ metric, which aggregates $\text{Row}_{\text{exec}}$ and $\text{Column}_{\text{exec}}$ to be the best indicator of performance.

As highlighted in the results, all components of the NL2KQL considerably contribute to the end-to-end performance.
Specifically, removing the schema (row 6), which originates from the Semantic Data Catalog, degrades $\text{Avg}_{\text{exec}}$ from 0.6350 down to 0.4309. 
Furthermore, excluding the few-shots from this approach (row 6), yields the solution on row 7, which shows a 46\% decrease in $\text{Avg}_{\text{exec}}$, bringing it to 0.2321. Conversely, adding the \textit{Schema Refiner} to the mix (row 5) improves this metric by 8.7\% from 0.5842.

Results show that removing \textit{Query Refiner} has a negative impact on seven out of eight metrics, with no change in the $\text{Table}$ metrics. This removal decreases the $\text{Avg}_{\text{exec}}$ by 1\% to 0.6265. 

For the \textit{Few-shot Selector} component, two alternatives were explored: the all-shots approach (including all few-shot examples)
and the zero-shot approach (removing few-shot examples). 
The results indicate that, in presence of the \textit{Schema Refiner} (rows 1, 2, 3, 4), the impact of few-shots saturate. Nonetheless, we still observe a 1\% improvement in the $\text{Avg}_{\text{exec}}$ metric from the all-shots solution (row 2) to the main proposed solution (row 1) with the addition of \textit{Few-shot Selector}. However, this addition results in a slight decrease of 0.2\% in both the $\text{Semantic}$ and $\text{Column}_{\text{exec}}$ metrics. We hypothesize that while the \textit{Few-shot Selector} enhances overall performance by providing targeted examples, it inadvertently introduces bias towards particular schema present in the few-shot examples. This bias can lead to a slight decrease in the model's generalization.
Despite NL2KQL consistently performed better across 7 out of the 8 metrics, the zero-shot achieved better results on the $\text{Table}$ metric. We hypothesize that the few-shots could have influenced the table selection in a way that negatively impacts this metric.

In the ablation study of \textit{Schema Refiner}, the entirety of the schema  was included in the model's context (row 5).
In another experiment the schema was excluded from the prompt (row 6). Both scenarios demonstrated subpar performance compared to the proposed NL2KQL, further showcasing the effectiveness of \textit{Schema Refiner}.

Finally, the NLQ of the evaluation set were directly asked from the GPT-4 model, 
which substantially under-performed compared to the NL2KQL solution, with $\text{Avg}_{\text{exec}}$ of 0.1150.
Furthermore, to highlight the impact of the instructions in the \textit{Prompt Builder}, the static sections of the NL2KQL prompt including KQL syntax guide and best practices were added to the prompt for GPT-4 which, on average, demonstrated $50\%$ better performance across all metrics.


\begin{table}[htbp]
\centering
\caption{Offline metrics. Comparison of the proposed method, NL2KQL, with baselines and ablated methods.}
\label{tab:main_results}
\resizebox{\linewidth}{!}{
\begin{tabular}{@{}lccccc@{}}
\toprule
Method & Syntax & Semantic & Table & $\text{Filter}_{\text{col}}$ & $\text{Filter}_{\text{lit}}$ \\
\midrule
NL2KQL \small{main}                    & \textbf{0.9878}	& 0.9603	& 0.8222	& \textbf{0.6992}	& \textbf{0.6656}         \\
NL2KQL \small{all-shots}             & 0.9846	& \textbf{0.9627}	& 0.8204	& 0.6965	& 0.6583 \\
NL2KQL \small{w/o Query Refiner}  & 0.9862	& 0.9400	& 0.8222	& 0.6979	& 0.6647 \\
NL2KQL \small{Zero-shot}          & 0.9830	& 0.9538	& \textbf{0.8414}	& 0.6600	& 0.6304 \\
NL2KQL \small{w/o Schema Refiner} & 0.9838	& 0.9213	& 0.8040	& 0.6735	& 0.6262 \\
NL2KQL \small{w/o Schema}         & 0.9854	& 0.6626	& 0.6765	& 0.5898	& 0.6334 \\
GPT-4 \small{with KQL Guide}         & 0.9586	& 0.3406	& 0.3731	& 0.4159	& 0.5610 \\
GPT-4 \small{w/o KQL Guide}             & 0.6513	& 0.1590	& 0.2238	& 0.3877	& 0.5639 \\ 
\bottomrule
\end{tabular}
}
\end{table}

\begin{table}[htbp]
\centering
\caption{Online metrics. Comparison of the proposed method, NL2KQL, with baselines and ablated methods.}
\label{tab:main_results_online}
\resizebox{\linewidth}{!}{
\tiny
\begin{tabular}{@{}lccc@{}}
\toprule
Method & $\text{Row}_{\text{exec}}$ & $\text{Column}_{\text{exec}}$  & $\text{Avg}_{\text{exec}}$ \\ 
\midrule
NL2KQL main    & \textbf{0.5335}	& 0.7365	& \textbf{0.6350}
        \\
NL2KQL all-shots             & 0.5147	& \textbf{0.7394}	& 0.6270
      \\
NL2KQL w/o Query Refiner  & 0.5270	& 0.7259	& 0.6265
      \\
NL2KQL Zero-shot          & 0.4989	& 0.6854	& 0.5922
       \\
NL2KQL w/o Schema Refiner & 0.4795	& 0.6888	& 0.5842
     \\
NL2KQL w/o Schema         & 0.3331	& 0.5287	& 0.4309
        \\
GPT-4 with KQL Guide         & 0.1727	& 0.2914	& 0.2321
        \\
GPT-4 w/o KQL Guide             & 0.0803	& 0.1497	& 0.1150
       \\
\bottomrule
\end{tabular}
}
\end{table}

\section{Discussion And Future Work}


In this work, we propose NL2KQL, an end-to-end Natural Language to Kusto Query Language translator which capitalizes on the capabilities of LLM. This system creates a dynamic prompt composed of the schema, few-shots, selective  elements of the Semantic Data Catalog, and details of the KQL syntax and its best practices to guide the model towards emitting the correct KQL. As demonstrated in ablation studies of Tables~\ref{tab:main_results} and \ref{tab:main_results_online}, the biggest boosts in performance of NL2KQL are rooted in the KQL Guide (syntax and best practices), Schema, and the Few-shots. 
Given the complexities of evaluating generated queries, we designed offline (based on parsing) and online (based on execution) metrics to capture the performance of the proposed method and compare it with available baselines. To the extent of our knowledge, this is the first NLQ-to-KQL solution.

While we had aimed to examine the performance of the proposed NL2KQL framework on other open-source and proprietary LLMs besides the GPT-4, however, 
to the extent of our knowledge, none of the other pre-trained LLMs included  KQL queries in their training data. 
The scope of this work does not cover training LLMs from scratch on KQL datasets.

The benchmark evaluation dataset, which accompanies and released along this paper, has only one KQL for each NLQ. 
However, 
considering that each NLQ can be addressed with  multiple equally valid KQLs, 
having a  single ground-truth KQL  can result in an under-estimation of the overall performance with respect to the offline metrics. 


Among future works, we envision the trending integration of LLMs as agent-like entities for iterative interaction, refinement and error resolution to be  a viable option. 
Given the real-time nature of NL2KQL, response time is a concern in design decisions.


\begin{acks}
The authors would like to thank Yogesh Roy, Sylvie Liu, Tali Ash, Liron Cohen, Inbar Rotem, Uday Kiran Ravuri, Vignesh Nayak, Jaimie Huang, Pete Bryan, Ram Shankar Siva Kumar, Lu Wang, Yi Mao, Weizhu Chen for their help in this work.
\end{acks}

\bibliographystyle{ACM-Reference-Format}
\bibliography{nl2kql}


\begin{thebibliography}{44}


\ifx \showCODEN    \undefined \def \showCODEN     #1{\unskip}     \fi
\ifx \showDOI      \undefined \def \showDOI       #1{#1}\fi
\ifx \showISBNx    \undefined \def \showISBNx     #1{\unskip}     \fi
\ifx \showISBNxiii \undefined \def \showISBNxiii  #1{\unskip}     \fi
\ifx \showISSN     \undefined \def \showISSN      #1{\unskip}     \fi
\ifx \showLCCN     \undefined \def \showLCCN      #1{\unskip}     \fi
\ifx \shownote     \undefined \def \shownote      #1{#1}          \fi
\ifx \showarticletitle \undefined \def \showarticletitle #1{#1}   \fi
\ifx \showURL      \undefined \def \showURL       {\relax}        \fi
\providecommand\bibfield[2]{#2}
\providecommand\bibinfo[2]{#2}
\providecommand\natexlab[1]{#1}
\providecommand\showeprint[2][]{arXiv:#2}

\bibitem[Ahmed and Devanbu(2023)]%
        {fewshot3}
\bibfield{author}{\bibinfo{person}{Toufique Ahmed} {and} \bibinfo{person}{Premkumar Devanbu}.} \bibinfo{year}{2023}\natexlab{}.
\newblock \showarticletitle{Few-shot training LLMs for project-specific code-summarization}. In \bibinfo{booktitle}{\emph{Proceedings of the 37th IEEE/ACM International Conference on Automated Software Engineering}} (<conf-loc>, <city>Rochester</city>, <state>MI</state>, <country>USA</country>, </conf-loc>) \emph{(\bibinfo{series}{ASE '22})}. \bibinfo{publisher}{Association for Computing Machinery}, \bibinfo{address}{New York, NY, USA}, Article \bibinfo{articleno}{177}, \bibinfo{numpages}{5}~pages.
\newblock
\showISBNx{9781450394758}
\urldef\tempurl%
\url{https://doi.org/10.1145/3551349.3559555}
\showDOI{\tempurl}


\bibitem[au2 et~al\mbox{.}(2021)]%
        {logan2021cutting}
\bibfield{author}{\bibinfo{person}{Robert L. Logan~IV au2}, \bibinfo{person}{Ivana Balažević}, \bibinfo{person}{Eric Wallace}, \bibinfo{person}{Fabio Petroni}, \bibinfo{person}{Sameer Singh}, {and} \bibinfo{person}{Sebastian Riedel}.} \bibinfo{year}{2021}\natexlab{}.
\newblock \bibinfo{title}{Cutting Down on Prompts and Parameters: Simple Few-Shot Learning with Language Models}.
\newblock
\newblock
\showeprint[arxiv]{2106.13353}~[cs.CL]


\bibitem[Baik et~al\mbox{.}(2019)]%
        {Christopher_2019}
\bibfield{author}{\bibinfo{person}{Christopher Baik}, \bibinfo{person}{H.~V. Jagadish}, {and} \bibinfo{person}{Yunyao Li}.} \bibinfo{year}{2019}\natexlab{}.
\newblock \showarticletitle{Bridging the Semantic Gap with SQL Query Logs in Natural Language Interfaces to Databases}. In \bibinfo{booktitle}{\emph{2019 IEEE 35th International Conference on Data Engineering (ICDE)}}. \bibinfo{publisher}{IEEE}.
\newblock
\urldef\tempurl%
\url{https://doi.org/10.1109/icde.2019.00041}
\showDOI{\tempurl}


\bibitem[Brown et~al\mbox{.}(2020a)]%
        {gpt2020}
\bibfield{author}{\bibinfo{person}{Tom Brown}, \bibinfo{person}{Benjamin Mann}, \bibinfo{person}{Nick Ryder}, \bibinfo{person}{Melanie Subbiah}, \bibinfo{person}{Jared~D Kaplan}, \bibinfo{person}{Prafulla Dhariwal}, \bibinfo{person}{Arvind Neelakantan}, \bibinfo{person}{Pranav Shyam}, \bibinfo{person}{Girish Sastry}, \bibinfo{person}{Amanda Askell}, \bibinfo{person}{Sandhini Agarwal}, \bibinfo{person}{Ariel Herbert-Voss}, \bibinfo{person}{Gretchen Krueger}, \bibinfo{person}{Tom Henighan}, \bibinfo{person}{Rewon Child}, \bibinfo{person}{Aditya Ramesh}, \bibinfo{person}{Daniel Ziegler}, \bibinfo{person}{Jeffrey Wu}, \bibinfo{person}{Clemens Winter}, \bibinfo{person}{Chris Hesse}, \bibinfo{person}{Mark Chen}, \bibinfo{person}{Eric Sigler}, \bibinfo{person}{Mateusz Litwin}, \bibinfo{person}{Scott Gray}, \bibinfo{person}{Benjamin Chess}, \bibinfo{person}{Jack Clark}, \bibinfo{person}{Christopher Berner}, \bibinfo{person}{Sam McCandlish}, \bibinfo{person}{Alec Radford}, \bibinfo{person}{Ilya Sutskever}, {and}
  \bibinfo{person}{Dario Amodei}.} \bibinfo{year}{2020}\natexlab{a}.
\newblock \showarticletitle{Language Models are Few-Shot Learners}. In \bibinfo{booktitle}{\emph{Advances in Neural Information Processing Systems}}, \bibfield{editor}{\bibinfo{person}{H.~Larochelle}, \bibinfo{person}{M.~Ranzato}, \bibinfo{person}{R.~Hadsell}, \bibinfo{person}{M.F. Balcan}, {and} \bibinfo{person}{H.~Lin}} (Eds.), Vol.~\bibinfo{volume}{33}. \bibinfo{publisher}{Curran Associates, Inc.}, \bibinfo{pages}{1877--1901}.
\newblock
\urldef\tempurl%
\url{https://proceedings.neurips.cc/paper_files/paper/2020/file/1457c0d6bfcb4967418bfb8ac142f64a-Paper.pdf}
\showURL{%
\tempurl}


\bibitem[Brown et~al\mbox{.}(2020b)]%
        {brown2020language}
\bibfield{author}{\bibinfo{person}{Tom~B. Brown}, \bibinfo{person}{Benjamin Mann}, \bibinfo{person}{Nick Ryder}, \bibinfo{person}{Melanie Subbiah}, \bibinfo{person}{Jared Kaplan}, \bibinfo{person}{Prafulla Dhariwal}, \bibinfo{person}{Arvind Neelakantan}, \bibinfo{person}{Pranav Shyam}, \bibinfo{person}{Girish Sastry}, \bibinfo{person}{Amanda Askell}, \bibinfo{person}{Sandhini Agarwal}, \bibinfo{person}{Ariel Herbert-Voss}, \bibinfo{person}{Gretchen Krueger}, \bibinfo{person}{Tom Henighan}, \bibinfo{person}{Rewon Child}, \bibinfo{person}{Aditya Ramesh}, \bibinfo{person}{Daniel~M. Ziegler}, \bibinfo{person}{Jeffrey Wu}, \bibinfo{person}{Clemens Winter}, \bibinfo{person}{Christopher Hesse}, \bibinfo{person}{Mark Chen}, \bibinfo{person}{Eric Sigler}, \bibinfo{person}{Mateusz Litwin}, \bibinfo{person}{Scott Gray}, \bibinfo{person}{Benjamin Chess}, \bibinfo{person}{Jack Clark}, \bibinfo{person}{Christopher Berner}, \bibinfo{person}{Sam McCandlish}, \bibinfo{person}{Alec Radford}, \bibinfo{person}{Ilya Sutskever},
  {and} \bibinfo{person}{Dario Amodei}.} \bibinfo{year}{2020}\natexlab{b}.
\newblock \bibinfo{title}{Language Models are Few-Shot Learners}.
\newblock
\newblock
\showeprint[arxiv]{2005.14165}~[cs.CL]


\bibitem[Brunner and Stockinger(2021)]%
        {brunner2021valuenet}
\bibfield{author}{\bibinfo{person}{Ursin Brunner} {and} \bibinfo{person}{Kurt Stockinger}.} \bibinfo{year}{2021}\natexlab{}.
\newblock \bibinfo{title}{ValueNet: A Natural Language-to-SQL System that Learns from Database Information}.
\newblock
\newblock
\showeprint[arxiv]{2006.00888}~[cs.DB]


\bibitem[Chen et~al\mbox{.}(2017)]%
        {chen-etal-2017-reading}
\bibfield{author}{\bibinfo{person}{Danqi Chen}, \bibinfo{person}{Adam Fisch}, \bibinfo{person}{Jason Weston}, {and} \bibinfo{person}{Antoine Bordes}.} \bibinfo{year}{2017}\natexlab{}.
\newblock \showarticletitle{Reading {W}ikipedia to Answer Open-Domain Questions}. In \bibinfo{booktitle}{\emph{Proceedings of the 55th Annual Meeting of the Association for Computational Linguistics (Volume 1: Long Papers)}}, \bibfield{editor}{\bibinfo{person}{Regina Barzilay} {and} \bibinfo{person}{Min-Yen Kan}} (Eds.). \bibinfo{publisher}{Association for Computational Linguistics}, \bibinfo{address}{Vancouver, Canada}, \bibinfo{pages}{1870--1879}.
\newblock
\urldef\tempurl%
\url{https://doi.org/10.18653/v1/P17-1171}
\showDOI{\tempurl}


\bibitem[Chen et~al\mbox{.}(2021)]%
        {codex}
\bibfield{author}{\bibinfo{person}{Mark Chen}, \bibinfo{person}{Jerry Tworek}, \bibinfo{person}{Heewoo Jun}, \bibinfo{person}{Qiming Yuan}, \bibinfo{person}{Henrique Ponde}, \bibinfo{person}{Jared Kaplan}, \bibinfo{person}{Harrison Edwards}, \bibinfo{person}{Yura Burda}, \bibinfo{person}{Nicholas Joseph}, \bibinfo{person}{Greg Brockman}, \bibinfo{person}{Alex Ray}, \bibinfo{person}{Raul Puri}, \bibinfo{person}{Gretchen Krueger}, \bibinfo{person}{Michael Petrov}, \bibinfo{person}{Heidy Khlaaf}, \bibinfo{person}{Girish Sastry}, \bibinfo{person}{Pamela Mishkin}, \bibinfo{person}{Brooke Chan}, \bibinfo{person}{Scott Gray}, \bibinfo{person}{Nick Ryder}, \bibinfo{person}{Mikhail Pavlov}, \bibinfo{person}{Alethea Power}, \bibinfo{person}{Lukasz Kaiser}, \bibinfo{person}{Mohammad Bavarian}, \bibinfo{person}{Clemens Winter}, \bibinfo{person}{Philippe Tillet}, \bibinfo{person}{Felipe~Petroski Such}, \bibinfo{person}{David~W. Cummings}, \bibinfo{person}{Matthias Plappert}, \bibinfo{person}{Fotios Chantzis},
  \bibinfo{person}{Elizabeth Barnes}, \bibinfo{person}{Ariel Herbert-Voss}, \bibinfo{person}{William~H. Guss}, \bibinfo{person}{Alex Nichol}, \bibinfo{person}{Igor Babuschkin}, \bibinfo{person}{Suchir Balaji}, \bibinfo{person}{Shantanu Jain}, \bibinfo{person}{Andrew Carr}, \bibinfo{person}{Jan Leike}, \bibinfo{person}{Joshua Achiam}, \bibinfo{person}{Vedant Misra}, \bibinfo{person}{Evan Morikawa}, \bibinfo{person}{Alec Radford}, \bibinfo{person}{Matthew~M. Knight}, \bibinfo{person}{Miles Brundage}, \bibinfo{person}{Mira Murati}, \bibinfo{person}{Katie Mayer}, \bibinfo{person}{Peter Welinder}, \bibinfo{person}{Bob McGrew}, \bibinfo{person}{Dario Amodei}, \bibinfo{person}{Sam McCandlish}, \bibinfo{person}{Ilya Sutskever}, {and} \bibinfo{person}{Wojciech Zaremba}.} \bibinfo{year}{2021}\natexlab{}.
\newblock \showarticletitle{Evaluating Large Language Models Trained on Code}.
\newblock \bibinfo{journal}{\emph{ArXiv}}  \bibinfo{volume}{abs/2107.03374} (\bibinfo{year}{2021}).
\newblock
\urldef\tempurl%
\url{https://api.semanticscholar.org/CorpusID:235755472}
\showURL{%
\tempurl}


\bibitem[Clark and Gardner(2018)]%
        {clark-gardner-2018-simple}
\bibfield{author}{\bibinfo{person}{Christopher Clark} {and} \bibinfo{person}{Matt Gardner}.} \bibinfo{year}{2018}\natexlab{}.
\newblock \showarticletitle{Simple and Effective Multi-Paragraph Reading Comprehension}. In \bibinfo{booktitle}{\emph{Proceedings of the 56th Annual Meeting of the Association for Computational Linguistics (Volume 1: Long Papers)}}, \bibfield{editor}{\bibinfo{person}{Iryna Gurevych} {and} \bibinfo{person}{Yusuke Miyao}} (Eds.). \bibinfo{publisher}{Association for Computational Linguistics}, \bibinfo{address}{Melbourne, Australia}, \bibinfo{pages}{845--855}.
\newblock
\urldef\tempurl%
\url{https://doi.org/10.18653/v1/P18-1078}
\showDOI{\tempurl}


\bibitem[Clement et~al\mbox{.}(2020)]%
        {Clement2020PyMT5MT}
\bibfield{author}{\bibinfo{person}{Colin~B. Clement}, \bibinfo{person}{Dawn Drain}, \bibinfo{person}{Jonathan Timcheck}, \bibinfo{person}{Alexey Svyatkovskiy}, {and} \bibinfo{person}{Neel Sundaresan}.} \bibinfo{year}{2020}\natexlab{}.
\newblock \showarticletitle{PyMT5: Multi-mode Translation of Natural Language and Python Code with Transformers}. In \bibinfo{booktitle}{\emph{Conference on Empirical Methods in Natural Language Processing}}.
\newblock
\urldef\tempurl%
\url{https://api.semanticscholar.org/CorpusID:222178041}
\showURL{%
\tempurl}


\bibitem[Dinan et~al\mbox{.}(2019)]%
        {dinan2018wizard}
\bibfield{author}{\bibinfo{person}{Emily Dinan}, \bibinfo{person}{Stephen Roller}, \bibinfo{person}{Kurt Shuster}, \bibinfo{person}{Angela Fan}, \bibinfo{person}{Michael Auli}, {and} \bibinfo{person}{Jason Weston}.} \bibinfo{year}{2019}\natexlab{}.
\newblock \showarticletitle{Wizard of Wikipedia: Knowledge-Powered Conversational Agents}. In \bibinfo{booktitle}{\emph{International Conference on Learning Representations}}.
\newblock
\urldef\tempurl%
\url{https://openreview.net/forum?id=r1l73iRqKm}
\showURL{%
\tempurl}


\bibitem[Feng et~al\mbox{.}(2020)]%
        {Feng2020CodeBERTAP}
\bibfield{author}{\bibinfo{person}{Zhangyin Feng}, \bibinfo{person}{Daya Guo}, \bibinfo{person}{Duyu Tang}, \bibinfo{person}{Nan Duan}, \bibinfo{person}{Xiaocheng Feng}, \bibinfo{person}{Ming Gong}, \bibinfo{person}{Linjun Shou}, \bibinfo{person}{Bing Qin}, \bibinfo{person}{Ting Liu}, \bibinfo{person}{Daxin Jiang}, {and} \bibinfo{person}{Ming Zhou}.} \bibinfo{year}{2020}\natexlab{}.
\newblock \showarticletitle{CodeBERT: A Pre-Trained Model for Programming and Natural Languages}.
\newblock \bibinfo{journal}{\emph{ArXiv}}  \bibinfo{volume}{abs/2002.08155} (\bibinfo{year}{2020}).
\newblock
\urldef\tempurl%
\url{https://api.semanticscholar.org/CorpusID:211171605}
\showURL{%
\tempurl}


\bibitem[Huang et~al\mbox{.}(2013)]%
        {inproceedings}
\bibfield{author}{\bibinfo{person}{Po-Sen Huang}, \bibinfo{person}{Xiaodong He}, \bibinfo{person}{Jianfeng Gao}, \bibinfo{person}{li Deng}, \bibinfo{person}{Alex Acero}, {and} \bibinfo{person}{Larry Heck}.} \bibinfo{year}{2013}\natexlab{}.
\newblock \showarticletitle{Learning deep structured semantic models for web search using clickthrough data}. \bibinfo{pages}{2333--2338}.
\newblock
\urldef\tempurl%
\url{https://doi.org/10.1145/2505515.2505665}
\showDOI{\tempurl}


\bibitem[Izacard and Grave(2022)]%
        {izacard2022distilling}
\bibfield{author}{\bibinfo{person}{Gautier Izacard} {and} \bibinfo{person}{Edouard Grave}.} \bibinfo{year}{2022}\natexlab{}.
\newblock \bibinfo{title}{Distilling Knowledge from Reader to Retriever for Question Answering}.
\newblock
\newblock
\showeprint[arxiv]{2012.04584}~[cs.CL]


\bibitem[Karpukhin et~al\mbox{.}(2020)]%
        {karpukhin2020dense}
\bibfield{author}{\bibinfo{person}{Vladimir Karpukhin}, \bibinfo{person}{Barlas Oğuz}, \bibinfo{person}{Sewon Min}, \bibinfo{person}{Patrick Lewis}, \bibinfo{person}{Ledell Wu}, \bibinfo{person}{Sergey Edunov}, \bibinfo{person}{Danqi Chen}, {and} \bibinfo{person}{Wen tau Yih}.} \bibinfo{year}{2020}\natexlab{}.
\newblock \bibinfo{title}{Dense Passage Retrieval for Open-Domain Question Answering}.
\newblock
\newblock
\showeprint[arxiv]{2004.04906}~[cs.CL]


\bibitem[Katsogiannis-Meimarakis and Koutrika(2023)]%
        {sqlsurvey}
\bibfield{author}{\bibinfo{person}{George Katsogiannis-Meimarakis} {and} \bibinfo{person}{Georgia Koutrika}.} \bibinfo{year}{2023}\natexlab{}.
\newblock \showarticletitle{A survey on deep learning approaches for text-to-SQL}.
\newblock \bibinfo{journal}{\emph{The VLDB Journal}}  \bibinfo{volume}{32} (\bibinfo{date}{01} \bibinfo{year}{2023}).
\newblock
\urldef\tempurl%
\url{https://doi.org/10.1007/s00778-022-00776-8}
\showDOI{\tempurl}


\bibitem[Kwiatkowski et~al\mbox{.}(2019)]%
        {kwiatkowski-etal-2019-natural}
\bibfield{author}{\bibinfo{person}{Tom Kwiatkowski}, \bibinfo{person}{Jennimaria Palomaki}, \bibinfo{person}{Olivia Redfield}, \bibinfo{person}{Michael Collins}, \bibinfo{person}{Ankur Parikh}, \bibinfo{person}{Chris Alberti}, \bibinfo{person}{Danielle Epstein}, \bibinfo{person}{Illia Polosukhin}, \bibinfo{person}{Jacob Devlin}, \bibinfo{person}{Kenton Lee}, \bibinfo{person}{Kristina Toutanova}, \bibinfo{person}{Llion Jones}, \bibinfo{person}{Matthew Kelcey}, \bibinfo{person}{Ming-Wei Chang}, \bibinfo{person}{Andrew~M. Dai}, \bibinfo{person}{Jakob Uszkoreit}, \bibinfo{person}{Quoc Le}, {and} \bibinfo{person}{Slav Petrov}.} \bibinfo{year}{2019}\natexlab{}.
\newblock \showarticletitle{Natural Questions: A Benchmark for Question Answering Research}.
\newblock \bibinfo{journal}{\emph{Transactions of the Association for Computational Linguistics}}  \bibinfo{volume}{7} (\bibinfo{year}{2019}), \bibinfo{pages}{452--466}.
\newblock
\urldef\tempurl%
\url{https://doi.org/10.1162/tacl_a_00276}
\showDOI{\tempurl}


\bibitem[Li and Jagadish(2014)]%
        {Fei10.14778/2735461.2735468}
\bibfield{author}{\bibinfo{person}{Fei Li} {and} \bibinfo{person}{H.~V. Jagadish}.} \bibinfo{year}{2014}\natexlab{}.
\newblock \showarticletitle{Constructing an interactive natural language interface for relational databases}.
\newblock \bibinfo{journal}{\emph{Proc. VLDB Endow.}} \bibinfo{volume}{8}, \bibinfo{number}{1} (\bibinfo{date}{sep} \bibinfo{year}{2014}), \bibinfo{pages}{73–84}.
\newblock
\showISSN{2150-8097}
\urldef\tempurl%
\url{https://doi.org/10.14778/2735461.2735468}
\showDOI{\tempurl}


\bibitem[Li et~al\mbox{.}(2023)]%
        {bird}
\bibfield{author}{\bibinfo{person}{Jinyang Li}, \bibinfo{person}{Binyuan Hui}, \bibinfo{person}{Ge Qu}, \bibinfo{person}{Jiaxi Yang}, \bibinfo{person}{Binhua Li}, \bibinfo{person}{Bowen Li}, \bibinfo{person}{Bailin Wang}, \bibinfo{person}{Bowen Qin}, \bibinfo{person}{Rongyu Cao}, \bibinfo{person}{Ruiying Geng}, \bibinfo{person}{Nan Huo}, \bibinfo{person}{Xuanhe Zhou}, \bibinfo{person}{Chenhao Ma}, \bibinfo{person}{Guoliang Li}, \bibinfo{person}{Kevin C.~C. Chang}, \bibinfo{person}{Fei Huang}, \bibinfo{person}{Reynold Cheng}, {and} \bibinfo{person}{Yongbin Li}.} \bibinfo{year}{2023}\natexlab{}.
\newblock \bibinfo{title}{Can LLM Already Serve as A Database Interface? A BIg Bench for Large-Scale Database Grounded Text-to-SQLs}.
\newblock
\newblock
\showeprint[arxiv]{2305.03111}~[cs.CL]


\bibitem[Radford et~al\mbox{.}(2019)]%
        {Radford2019LanguageMA}
\bibfield{author}{\bibinfo{person}{Alec Radford}, \bibinfo{person}{Jeff Wu}, \bibinfo{person}{Rewon Child}, \bibinfo{person}{David Luan}, \bibinfo{person}{Dario Amodei}, {and} \bibinfo{person}{Ilya Sutskever}.} \bibinfo{year}{2019}\natexlab{}.
\newblock \showarticletitle{Language Models are Unsupervised Multitask Learners}.
\newblock
\urldef\tempurl%
\url{https://api.semanticscholar.org/CorpusID:160025533}
\showURL{%
\tempurl}


\bibitem[Saha et~al\mbox{.}(2016)]%
        {Diptikalyan10.14778/2994509.2994536}
\bibfield{author}{\bibinfo{person}{Diptikalyan Saha}, \bibinfo{person}{Avrilia Floratou}, \bibinfo{person}{Karthik Sankaranarayanan}, \bibinfo{person}{Umar~Farooq Minhas}, \bibinfo{person}{Ashish~R. Mittal}, {and} \bibinfo{person}{Fatma \"{O}zcan}.} \bibinfo{year}{2016}\natexlab{}.
\newblock \showarticletitle{ATHENA: an ontology-driven system for natural language querying over relational data stores}.
\newblock \bibinfo{journal}{\emph{Proc. VLDB Endow.}} \bibinfo{volume}{9}, \bibinfo{number}{12} (\bibinfo{date}{aug} \bibinfo{year}{2016}), \bibinfo{pages}{1209–1220}.
\newblock
\showISSN{2150-8097}
\urldef\tempurl%
\url{https://doi.org/10.14778/2994509.2994536}
\showDOI{\tempurl}


\bibitem[Scholak et~al\mbox{.}(2021)]%
        {Torsten}
\bibfield{author}{\bibinfo{person}{Torsten Scholak}, \bibinfo{person}{Nathan Schucher}, {and} \bibinfo{person}{Dzmitry Bahdanau}.} \bibinfo{year}{2021}\natexlab{}.
\newblock \bibinfo{title}{PICARD: Parsing Incrementally for Constrained Auto-Regressive Decoding from Language Models}.
\newblock
\newblock
\showeprint[arxiv]{2109.05093}~[cs.CL]


\bibitem[Sen et~al\mbox{.}(2020)]%
        {Jaydeep2020ATHENANL}
\bibfield{author}{\bibinfo{person}{Jaydeep Sen}, \bibinfo{person}{Chuan Lei}, \bibinfo{person}{Abdul Quamar}, \bibinfo{person}{Fatma {\"O}zcan}, \bibinfo{person}{Vasilis Efthymiou}, \bibinfo{person}{Ayushi Dalmia}, \bibinfo{person}{Greg Stager}, \bibinfo{person}{Ashish~R. Mittal}, \bibinfo{person}{Diptikalyan Saha}, {and} \bibinfo{person}{Karthik Sankaranarayanan}.} \bibinfo{year}{2020}\natexlab{}.
\newblock \showarticletitle{ATHENA++: Natural Language Querying for Complex Nested SQL Queries}.
\newblock \bibinfo{journal}{\emph{Proc. VLDB Endow.}}  \bibinfo{volume}{13} (\bibinfo{year}{2020}), \bibinfo{pages}{2747--2759}.
\newblock
\urldef\tempurl%
\url{https://api.semanticscholar.org/CorpusID:221348677}
\showURL{%
\tempurl}


\bibitem[Shen et~al\mbox{.}(2014)]%
        {10.1145/2567948.2577348}
\bibfield{author}{\bibinfo{person}{Yelong Shen}, \bibinfo{person}{Xiaodong He}, \bibinfo{person}{Jianfeng Gao}, \bibinfo{person}{Li Deng}, {and} \bibinfo{person}{Gr\'{e}goire Mesnil}.} \bibinfo{year}{2014}\natexlab{}.
\newblock \showarticletitle{Learning semantic representations using convolutional neural networks for web search}. In \bibinfo{booktitle}{\emph{Proceedings of the 23rd International Conference on World Wide Web}} (Seoul, Korea) \emph{(\bibinfo{series}{WWW '14 Companion})}. \bibinfo{publisher}{Association for Computing Machinery}, \bibinfo{address}{New York, NY, USA}, \bibinfo{pages}{373–374}.
\newblock
\showISBNx{9781450327459}
\urldef\tempurl%
\url{https://doi.org/10.1145/2567948.2577348}
\showDOI{\tempurl}


\bibitem[Shuster et~al\mbox{.}(2021)]%
        {shuster2021retrieval}
\bibfield{author}{\bibinfo{person}{Kurt Shuster}, \bibinfo{person}{Spencer Poff}, \bibinfo{person}{Moya Chen}, \bibinfo{person}{Douwe Kiela}, {and} \bibinfo{person}{Jason Weston}.} \bibinfo{year}{2021}\natexlab{}.
\newblock \bibinfo{title}{Retrieval Augmentation Reduces Hallucination in Conversation}.
\newblock
\newblock
\showeprint[arxiv]{2104.07567}~[cs.CL]


\bibitem[Simon(1963)]%
        {simon63}
\bibfield{author}{\bibinfo{person}{Herbert~A. Simon}.} \bibinfo{year}{1963}\natexlab{}.
\newblock \showarticletitle{Experiments with a Heuristic Compiler}.
\newblock \bibinfo{journal}{\emph{J. ACM}} \bibinfo{volume}{10}, \bibinfo{number}{4} (\bibinfo{date}{oct} \bibinfo{year}{1963}), \bibinfo{pages}{493–506}.
\newblock
\showISSN{0004-5411}
\urldef\tempurl%
\url{https://doi.org/10.1145/321186.321192}
\showDOI{\tempurl}


\bibitem[Summers(1977)]%
        {summers77}
\bibfield{author}{\bibinfo{person}{Phillip~D. Summers}.} \bibinfo{year}{1977}\natexlab{}.
\newblock \showarticletitle{A Methodology for LISP Program Construction from Examples}.
\newblock \bibinfo{journal}{\emph{J. ACM}} \bibinfo{volume}{24}, \bibinfo{number}{1} (\bibinfo{date}{jan} \bibinfo{year}{1977}), \bibinfo{pages}{161–175}.
\newblock
\showISSN{0004-5411}
\urldef\tempurl%
\url{https://doi.org/10.1145/321992.322002}
\showDOI{\tempurl}


\bibitem[Thorne et~al\mbox{.}(2018)]%
        {thorne2018fever}
\bibfield{author}{\bibinfo{person}{James Thorne}, \bibinfo{person}{Andreas Vlachos}, \bibinfo{person}{Christos Christodoulopoulos}, {and} \bibinfo{person}{Arpit Mittal}.} \bibinfo{year}{2018}\natexlab{}.
\newblock \bibinfo{title}{FEVER: a large-scale dataset for Fact Extraction and VERification}.
\newblock
\newblock
\showeprint[arxiv]{1803.05355}~[cs.CL]


\bibitem[Vinyals et~al\mbox{.}(2017)]%
        {vinyals2017matching}
\bibfield{author}{\bibinfo{person}{Oriol Vinyals}, \bibinfo{person}{Charles Blundell}, \bibinfo{person}{Timothy Lillicrap}, \bibinfo{person}{Koray Kavukcuoglu}, {and} \bibinfo{person}{Daan Wierstra}.} \bibinfo{year}{2017}\natexlab{}.
\newblock \bibinfo{title}{Matching Networks for One Shot Learning}.
\newblock
\newblock
\showeprint[arxiv]{1606.04080}~[cs.LG]


\bibitem[Voorhees and Tice(2000)]%
        {voorhees-tice-2000-trec}
\bibfield{author}{\bibinfo{person}{Ellen~M. Voorhees} {and} \bibinfo{person}{Dawn~M. Tice}.} \bibinfo{year}{2000}\natexlab{}.
\newblock \showarticletitle{The {TREC}-8 Question Answering Track}. In \bibinfo{booktitle}{\emph{Proceedings of the Second International Conference on Language Resources and Evaluation ({LREC}{'}00)}}, \bibfield{editor}{\bibinfo{person}{M.~Gavrilidou}, \bibinfo{person}{G.~Carayannis}, \bibinfo{person}{S.~Markantonatou}, \bibinfo{person}{S.~Piperidis}, {and} \bibinfo{person}{G.~Stainhauer}} (Eds.). \bibinfo{publisher}{European Language Resources Association (ELRA)}, \bibinfo{address}{Athens, Greece}.
\newblock
\urldef\tempurl%
\url{http://www.lrec-conf.org/proceedings/lrec2000/pdf/26.pdf}
\showURL{%
\tempurl}


\bibitem[Waldinger and Lee(1969)]%
        {waldinger69}
\bibfield{author}{\bibinfo{person}{Richard~J. Waldinger} {and} \bibinfo{person}{Richard C.~T. Lee}.} \bibinfo{year}{1969}\natexlab{}.
\newblock \showarticletitle{PROW: a step toward automatic program writing}. In \bibinfo{booktitle}{\emph{Proceedings of the 1st International Joint Conference on Artificial Intelligence}} (Washington, DC) \emph{(\bibinfo{series}{IJCAI'69})}. \bibinfo{publisher}{Morgan Kaufmann Publishers Inc.}, \bibinfo{address}{San Francisco, CA, USA}, \bibinfo{pages}{241–252}.
\newblock


\bibitem[Wang et~al\mbox{.}(2020)]%
        {Bailin-etal-2020-rat}
\bibfield{author}{\bibinfo{person}{Bailin Wang}, \bibinfo{person}{Richard Shin}, \bibinfo{person}{Xiaodong Liu}, \bibinfo{person}{Oleksandr Polozov}, {and} \bibinfo{person}{Matthew Richardson}.} \bibinfo{year}{2020}\natexlab{}.
\newblock \showarticletitle{{RAT-SQL}: Relation-Aware Schema Encoding and Linking for Text-to-{SQL} Parsers}. In \bibinfo{booktitle}{\emph{Proceedings of the 58th Annual Meeting of the Association for Computational Linguistics}}, \bibfield{editor}{\bibinfo{person}{Dan Jurafsky}, \bibinfo{person}{Joyce Chai}, \bibinfo{person}{Natalie Schluter}, {and} \bibinfo{person}{Joel Tetreault}} (Eds.). \bibinfo{publisher}{Association for Computational Linguistics}, \bibinfo{address}{Online}, \bibinfo{pages}{7567--7578}.
\newblock
\urldef\tempurl%
\url{https://doi.org/10.18653/v1/2020.acl-main.677}
\showDOI{\tempurl}


\bibitem[Wang et~al\mbox{.}(2019)]%
        {wang2019multipassage}
\bibfield{author}{\bibinfo{person}{Zhiguo Wang}, \bibinfo{person}{Patrick Ng}, \bibinfo{person}{Xiaofei Ma}, \bibinfo{person}{Ramesh Nallapati}, {and} \bibinfo{person}{Bing Xiang}.} \bibinfo{year}{2019}\natexlab{}.
\newblock \bibinfo{title}{Multi-passage BERT: A Globally Normalized BERT Model for Open-domain Question Answering}.
\newblock
\newblock
\showeprint[arxiv]{1908.08167}~[cs.CL]


\bibitem[Wei et~al\mbox{.}(2022a)]%
        {wei2022emergent}
\bibfield{author}{\bibinfo{person}{Jason Wei}, \bibinfo{person}{Yi Tay}, \bibinfo{person}{Rishi Bommasani}, \bibinfo{person}{Colin Raffel}, \bibinfo{person}{Barret Zoph}, \bibinfo{person}{Sebastian Borgeaud}, \bibinfo{person}{Dani Yogatama}, \bibinfo{person}{Maarten Bosma}, \bibinfo{person}{Denny Zhou}, \bibinfo{person}{Donald Metzler}, \bibinfo{person}{Ed~H. Chi}, \bibinfo{person}{Tatsunori Hashimoto}, \bibinfo{person}{Oriol Vinyals}, \bibinfo{person}{Percy Liang}, \bibinfo{person}{Jeff Dean}, {and} \bibinfo{person}{William Fedus}.} \bibinfo{year}{2022}\natexlab{a}.
\newblock \bibinfo{title}{Emergent Abilities of Large Language Models}.
\newblock
\newblock
\showeprint[arxiv]{2206.07682}~[cs.CL]


\bibitem[Wei et~al\mbox{.}(2022b)]%
        {CoT}
\bibfield{author}{\bibinfo{person}{Jason Wei}, \bibinfo{person}{Xuezhi Wang}, \bibinfo{person}{Dale Schuurmans}, \bibinfo{person}{Maarten Bosma}, \bibinfo{person}{Ed~Huai hsin Chi}, \bibinfo{person}{F. Xia}, \bibinfo{person}{Quoc Le}, {and} \bibinfo{person}{Denny Zhou}.} \bibinfo{year}{2022}\natexlab{b}.
\newblock \showarticletitle{Chain of Thought Prompting Elicits Reasoning in Large Language Models}.
\newblock \bibinfo{journal}{\emph{ArXiv}}  \bibinfo{volume}{abs/2201.11903} (\bibinfo{year}{2022}).
\newblock
\urldef\tempurl%
\url{https://api.semanticscholar.org/CorpusID:246411621}
\showURL{%
\tempurl}


\bibitem[Xiong et~al\mbox{.}(2020)]%
        {xiong2020approximate}
\bibfield{author}{\bibinfo{person}{Lee Xiong}, \bibinfo{person}{Chenyan Xiong}, \bibinfo{person}{Ye Li}, \bibinfo{person}{Kwok-Fung Tang}, \bibinfo{person}{Jialin Liu}, \bibinfo{person}{Paul Bennett}, \bibinfo{person}{Junaid Ahmed}, {and} \bibinfo{person}{Arnold Overwijk}.} \bibinfo{year}{2020}\natexlab{}.
\newblock \bibinfo{title}{Approximate Nearest Neighbor Negative Contrastive Learning for Dense Text Retrieval}.
\newblock
\newblock
\showeprint[arxiv]{2007.00808}~[cs.IR]


\bibitem[Xu et~al\mbox{.}(2017)]%
        {Xiaojun}
\bibfield{author}{\bibinfo{person}{Xiaojun Xu}, \bibinfo{person}{Chang Liu}, {and} \bibinfo{person}{Dawn Song}.} \bibinfo{year}{2017}\natexlab{}.
\newblock \bibinfo{title}{SQLNet: Generating Structured Queries From Natural Language Without Reinforcement Learning}.
\newblock
\newblock
\showeprint[arxiv]{1711.04436}~[cs.CL]


\bibitem[Yates et~al\mbox{.}(2021)]%
        {10.1145/3437963.3441667}
\bibfield{author}{\bibinfo{person}{Andrew Yates}, \bibinfo{person}{Rodrigo Nogueira}, {and} \bibinfo{person}{Jimmy Lin}.} \bibinfo{year}{2021}\natexlab{}.
\newblock \showarticletitle{Pretrained Transformers for Text Ranking: BERT and Beyond}. In \bibinfo{booktitle}{\emph{Proceedings of the 14th ACM International Conference on Web Search and Data Mining}} (Virtual Event, Israel) \emph{(\bibinfo{series}{WSDM '21})}. \bibinfo{publisher}{Association for Computing Machinery}, \bibinfo{address}{New York, NY, USA}, \bibinfo{pages}{1154–1156}.
\newblock
\showISBNx{9781450382977}
\urldef\tempurl%
\url{https://doi.org/10.1145/3437963.3441667}
\showDOI{\tempurl}


\bibitem[Yih et~al\mbox{.}(2011)]%
        {yih-etal-2011-learning}
\bibfield{author}{\bibinfo{person}{Wen-tau Yih}, \bibinfo{person}{Kristina Toutanova}, \bibinfo{person}{John~C. Platt}, {and} \bibinfo{person}{Christopher Meek}.} \bibinfo{year}{2011}\natexlab{}.
\newblock \showarticletitle{Learning Discriminative Projections for Text Similarity Measures}. In \bibinfo{booktitle}{\emph{Proceedings of the Fifteenth Conference on Computational Natural Language Learning}}, \bibfield{editor}{\bibinfo{person}{Sharon Goldwater} {and} \bibinfo{person}{Christopher Manning}} (Eds.). \bibinfo{publisher}{Association for Computational Linguistics}, \bibinfo{address}{Portland, Oregon, USA}, \bibinfo{pages}{247--256}.
\newblock
\urldef\tempurl%
\url{https://aclanthology.org/W11-0329}
\showURL{%
\tempurl}


\bibitem[Yu et~al\mbox{.}(2018a)]%
        {Tao}
\bibfield{author}{\bibinfo{person}{Tao Yu}, \bibinfo{person}{Michihiro Yasunaga}, \bibinfo{person}{Kai Yang}, \bibinfo{person}{Rui Zhang}, \bibinfo{person}{Dongxu Wang}, \bibinfo{person}{Zifan Li}, {and} \bibinfo{person}{Dragomir Radev}.} \bibinfo{year}{2018}\natexlab{a}.
\newblock \bibinfo{title}{SyntaxSQLNet: Syntax Tree Networks for Complex and Cross-DomainText-to-SQL Task}.
\newblock
\newblock
\showeprint[arxiv]{1810.05237}~[cs.CL]


\bibitem[Yu et~al\mbox{.}(2018b)]%
        {spider}
\bibfield{author}{\bibinfo{person}{Tao Yu}, \bibinfo{person}{Rui Zhang}, \bibinfo{person}{Kai Yang}, \bibinfo{person}{Michihiro Yasunaga}, \bibinfo{person}{Dongxu Wang}, \bibinfo{person}{Zifan Li}, \bibinfo{person}{James Ma}, \bibinfo{person}{Irene Li}, \bibinfo{person}{Qingning Yao}, \bibinfo{person}{Shanelle Roman}, \bibinfo{person}{Zilin Zhang}, {and} \bibinfo{person}{Dragomir Radev}.} \bibinfo{year}{2018}\natexlab{b}.
\newblock \showarticletitle{{S}pider: A Large-Scale Human-Labeled Dataset for Complex and Cross-Domain Semantic Parsing and Text-to-{SQL} Task}. In \bibinfo{booktitle}{\emph{Proceedings of the 2018 Conference on Empirical Methods in Natural Language Processing}}, \bibfield{editor}{\bibinfo{person}{Ellen Riloff}, \bibinfo{person}{David Chiang}, \bibinfo{person}{Julia Hockenmaier}, {and} \bibinfo{person}{Jun{'}ichi Tsujii}} (Eds.). \bibinfo{publisher}{Association for Computational Linguistics}, \bibinfo{address}{Brussels, Belgium}, \bibinfo{pages}{3911--3921}.
\newblock
\urldef\tempurl%
\url{https://doi.org/10.18653/v1/D18-1425}
\showDOI{\tempurl}


\bibitem[Zhang et~al\mbox{.}(2024)]%
        {text2sqlbenchmark}
\bibfield{author}{\bibinfo{person}{Bin Zhang}, \bibinfo{person}{Yuxiao Ye}, \bibinfo{person}{Guoqing Du}, \bibinfo{person}{Xiaoru Hu}, \bibinfo{person}{Zhishuai Li}, \bibinfo{person}{Sun Yang}, \bibinfo{person}{Chi~Harold Liu}, \bibinfo{person}{Rui Zhao}, \bibinfo{person}{Ziyue Li}, {and} \bibinfo{person}{Hangyu Mao}.} \bibinfo{year}{2024}\natexlab{}.
\newblock \bibinfo{title}{Benchmarking the Text-to-SQL Capability of Large Language Models: A Comprehensive Evaluation}.
\newblock
\newblock
\showeprint[arxiv]{2403.02951}~[cs.CL]


\bibitem[Zhong et~al\mbox{.}(2017)]%
        {wikisql}
\bibfield{author}{\bibinfo{person}{Victor Zhong}, \bibinfo{person}{Caiming Xiong}, {and} \bibinfo{person}{Richard Socher}.} \bibinfo{year}{2017}\natexlab{}.
\newblock \showarticletitle{Seq2SQL: Generating Structured Queries from Natural Language using Reinforcement Learning}.
\newblock \bibinfo{journal}{\emph{CoRR}}  \bibinfo{volume}{abs/1709.00103} (\bibinfo{year}{2017}).
\newblock


\bibitem[Ziegler et~al\mbox{.}(2022)]%
        {productivity22}
\bibfield{author}{\bibinfo{person}{Albert Ziegler}, \bibinfo{person}{Eirini Kalliamvakou}, \bibinfo{person}{X.~Alice Li}, \bibinfo{person}{Andrew Rice}, \bibinfo{person}{Devon Rifkin}, \bibinfo{person}{Shawn Simister}, \bibinfo{person}{Ganesh Sittampalam}, {and} \bibinfo{person}{Edward Aftandilian}.} \bibinfo{year}{2022}\natexlab{}.
\newblock \showarticletitle{Productivity assessment of neural code completion}. In \bibinfo{booktitle}{\emph{Proceedings of the 6th ACM SIGPLAN International Symposium on Machine Programming}} (San Diego, CA, USA) \emph{(\bibinfo{series}{MAPS 2022})}. \bibinfo{publisher}{Association for Computing Machinery}, \bibinfo{address}{New York, NY, USA}, \bibinfo{pages}{21–29}.
\newblock
\showISBNx{9781450392730}
\urldef\tempurl%
\url{https://doi.org/10.1145/3520312.3534864}
\showDOI{\tempurl}


\end{thebibliography}

\appendix

\section{Prompt Template}
\label{apx:prompt-template}

Each section of the prompt fill its respective \lstinline|{{NAME_PLACEHOLDER}}|  placeholder in the following prompt template:

\begin{lstlisting} [basicstyle=\small, frame=single, breaklines=true, columns=fullflexible]
# Instructions

You will receive a natural language request. As a cyber-security expert specializing in Microsoft USX, your task is to write a Kusto Query Language (KQL) query that fulfills the request while adhering to all best practices.

Follow these steps:
- Start by coming up with a stepwise plan on how to address the request.
- Identify any challenges you might face with this plan.
- Choose a final approach that seems to fit the request best.
- Write an optimized and performant Kusto query that fulfills the request while adhering to all best practices.

# Schema
    
To help you write your query, you have been given the following Microsoft USX database schema. 
Use your own discretion to decide which tables and columns are needed.

{{SCHEMA_PLACEHOLDER}}

# Possible Values
In addition, the user has identified the following values which can appear in these tables.
Use your own discretion whether to include any of these values and how best to use them.
Do not use any values that are not explicitly mentioned in the prompt or the section below.

{{VALUES_PLACEHOLDER}}

# Kusto Syntax

A sample of useful KQL syntax are listed below.

**Scalar Functions**

| Syntax | Usage |
| ------ | ----- |
| `around(value, center, delta)` | Check if `value` is in the range `center +/- delta` |
| `bin(value, roundto)` | Round `value` to the nearest multiple of `roundto` |
| `isempty(value)` | Check if a column is null or empty |
| `parse_json(string)` | Convert a JSON-like string into a `dynamic` object |
| `todouble(value)` | Convert a value to a double |
| `toint(value)` | Convert a value to an integer |
| `toscalar(expression)` | Convert a single-row, single-column tabular expression into a scalar value |
| `tostring(value)` | Convert a value to a string |
| `range ( start, stop[, step ])` | Create a dynamic array of equally spaced values |

**Aggregation Functions**

| Syntax | Usage |
| ------ | ----- |
| `arg_min(minExpr, returnExpr [, ...])` | Find the row that minimizes an expression |
| `count()` | Count the number of records in a table or group |
| `count_distinct(expr)` | Count the number of unique values of `expr` per group |
| `dcount(expr)` | Estimate the number of unique values of `expr` per group |
| `make_set(expr)` | Create an array of unique values of `expr` per group |
| `percentile(expr, percentile)` | Estimate the nearest-rank percentile of the population defined by `expr` |

**Window Functions**

| Syntax | Usage |
| ------ | ----- |
| `row_number()` | Get the index of each row |
| `next(column[, offset, default_value ])` | Get the value of an upcoming row |
| `prev(column[, offset, default_value ])` | Get the value of a previous row |

**Tabular Operators**

| Syntax | Usage |
| ------ | ----- |
| `distinct ColumnName[, ColumnName2, ...]` | Produce a table of unique combinations of the input columns |
| `mv-apply Columns [to typeof(DataType)] on ( SubQuery )` | Apply a subquery to each record and union the results |
| `mv-expand [bagexpansion=(bag | array)] [Name =] Expr [to typeof(DataType)]` | Explode an array or JSON object into multiple rows |
| `render Visualization` | Render a visualization |
| `serialize` | Enable window functions |

**Scalar Operators**

| Syntax | Usage |
| ------ | ----- |
| `contains ( string )` | Check if a value contains a substring |
| `has ( string )` | Check if a value contains a specific word or term |
| `has_any ( string, ... )` | Check if a value contains any substring in a set |
| `has_all ( string, ... )` | Check if a value contains all substrings in a set |
| `in~ ( string, ... )` | Check if a value equals any substring in a set |
| `in ( number, ... )` | Check if a value equals any number in a set |
| `matches regex string` | Check if a value matches a regex pattern |

# Kusto Best practices

**General Guidelines**

- Never use placeholder values
- Minimize the number of tables and columns referenced
- Only include user-provided values in your queries with the correct table and column
- All backslashes in strings must be properly escaped
- Columns can be renamed using the `=` operator
- Avoid multiple filter conditions in a single `where` statement
- Functions that create columns will typically be named like `function_`, `function_column`, etc.

**Operators**

- Use `has_all`, `has_any`, `in~`, and `in` in place of multiple `and` or `or` operators
- Consider using `has` or `contains` before parsing JSON columns to avoid expensive parsing of rows without the required keys or values
- Use case-insensitive operators when comparing string and dynamic columns

**Aggregations**

- Aggregation functions only appear in summarize statements
- When combining aggregations and scalar values, a self-join may be necessary
- \"arg\" functions do not rename columns
- Aggregation functions cannot be nested. Use two separate summarize statements instead e.g., `| summarize X=AggFunc1(Col1) by Col2 | summarize AggFunc2(X)`

**Joining Tables**

- `join` conditions **MUST** contain only `==` 
- Columns that appear in multiple tables merged by a `join` statement have integer suffixes e.g., ColumnName, ColumnName1, etc
- Consider all kinds of join. Use 'innerunique' to keep all columns from both tables
- Use semi-joins (`leftsemi`, `leftantisemi`, `rightsemi`, etc.) to only keep columns from one table
- Use anti-joins (`leftanti`, `leftantisemi`, `rightanti`, etc.) to exclude records that appear in a table from another

**Tips and Tricks**

- Inequality operators like `between`, `<=`, etc. cannot be used to join tables. To join on a window on datetimes or other high-cardinality columns, use `bin` and `range` to define a new key
~~~kusto
Table1 
| extend Key=bin(TimeColumn, WindowSize) 
| join ( 
    Table2 
    | mv-expand Key=range(bin(TimeColumn - WindowSize, WindowSize / 2), bin(TimeColumn, WindowSize / 2), WindowSize / 2) to typeof(datetime) 
) on Key 
| where TimeColumn1 - TimeColumn between (0m .. WindowSize)
~~~

- To compare an aggregated and unaggregated version of the same column, you will need to use a self-join
~~~kusto
Table
| summarize VarName=AggFunc(Col1) by Col2 
| join Table on $left.Col2 == $right.Col2 
| where Col1 < VarName
~~~

- Some columns may be blank or dependent on other column values. Use a self-merge to utilize more than one of these columns at a time
~~~kusto
Table
| where IndependentColumn == Value1
| where isnotempty(DependentColumn1)
| join (
    Table 
    | where IndependentColumn == Value2
    | where isnotempty(DependentColumn2)
) on JoinColumn
| project JoinColumn, IndependentColumn, DependentColumn1, DependentColumn2
~~~

# Examples

{{EXAMPLES_PLACEHOLDER}}

# Reminder

All steps must be completed in a single message.
Remember, your output will contain queries like
~~~kusto
KQL QUERY GOES HERE
~~~

{{USER_REQUEST_PLACEHOLDER}}

\end{lstlisting}

\section{KQL Forms}
\label{apx:kql-forms}

KQL has a pipeline nature where data flows through a series of transformations separated with the pipe $|$ tokens.
Not all KQL queries share the same structure, but a typical KQL query fits in the following template:

\begin{lstlisting} [basicstyle=\small, frame=single, breaklines=true, columns=fullflexible]
<TableExpression>
| <operator> [parameters]
| <operator> [parameters]
...
\end{lstlisting}
where \lstinline{TableExpression} specifies the data source as a single table name identifier, joined tables, or a subquery that produces a table.
Operators transform the data including filtering (\lstinline{where}), projecting specific fields (\lstinline{project}), aggregating (\lstinline{summarize}), sorting (\lstinline{order}), and more.
Some simple KQL queries which capture generic basic use-cases are listed here.\\
\begin{lstlisting} [basicstyle=\small, frame=single, breaklines=true, columns=fullflexible]
// Data Retrieval
TableName 
| take 100

// Filtering
TableName 
| where ColName == 'Value'

// Aggregation
TableName 
| summarize Count = count() by ColName

// Sorting
TableName 
| order by ColName asc
\end{lstlisting}
The conventional pipeline structure in KQL (TABLE | OPERATOR PARAMETER) is not strictly followed by all KQL queries. Here are a few KQL syntax which do not fit int the pipeline structure.

\subsection{Let Statements}
KQL allows the definition of variables using the let statement, which can be used to simplify complex queries, store intermediate results, or define constants. The let statements do not fit the conventional pipeline structure as it precedes the main query.

\begin{lstlisting} [basicstyle=\small, frame=single, breaklines=true, columns=fullflexible]
let Threshold = 100;
T | where Value > Threshold    
\end{lstlisting}

\subsection{Subqueries}
Subqueries can be used within a larger query and do not always follow the linear pipeline structure. They can appear in various contexts, such as within a project or extend operator, or as part of a join condition.

\begin{lstlisting} [basicstyle=\small, frame=single, breaklines=true, columns=fullflexible]
T 
| extend NewValue = (subquery 
| summarize avg(Value))
\end{lstlisting}

\subsection{Scalar and Tabular Functions}
KQL supports both scalar and tabular functions. Scalar functions can be used within expressions and do not follow the pipeline pattern. Tabular functions, on the other hand, can generate tables that may be used at the beginning of a query or within a subquery.

\begin{lstlisting} [basicstyle=\small, frame=single, breaklines=true, columns=fullflexible]
// Scalar function example
T | where strlen(Name) > 10

// Tabular function example
range x from 1 to 10 step 1 
| summarize sum(x)
\end{lstlisting}

\subsection{Control Commands}
Control commands in KQL are used for management and metadata operations, such as creating tables or altering table schemas. These commands do not operate on data in the same way as the pipeline operators and do not follow the pipeline structure. Here is an example of a $create$ command.

\begin{lstlisting} [basicstyle=\small, frame=single, breaklines=true, columns=fullflexible]
.create table MyTable (Name: string, Value: int)
\end{lstlisting}

\subsection{Union Operator}
The union operator is used to combine rows from two or more tables or queries. While it can be part of a pipeline, its usage diverges from the simple  transformation sequence of pipes as it merges multiple data sources.

\begin{lstlisting} [basicstyle=\small, frame=single, breaklines=true, columns=fullflexible]
union T1, T2 
| where Value > 50
\end{lstlisting}

\subsection{Conditional Statements}
KQL supports conditional logic within queries, such as the case statement, which does not fit the  pipeline model but is embedded within expressions.
\begin{lstlisting} [basicstyle=\small, frame=single, breaklines=true, columns=fullflexible]
T 
| extend Status = case (
    Value > 100, "High", 
    Value < 50, "Low", 
    "Medium"
)
\end{lstlisting}

\subsection{JSON Parse and Join}
\label{sec:parse-json-join}
It is a common pattern in KQL to parse a JSON column and use its values to join tables.

\begin{lstlisting} [basicstyle=\small, frame=single, breaklines=true, columns=fullflexible]
TableA
| extend JoinKey = tostring(parse_json(JsonColumn).JoinKey)
| join kind=inner TableB on $left.JoinKey == $right.KeyColumn
\end{lstlisting}

\section{Database Schema}
\label{apx:schema}


\subsection{Defender Database}
\begin{lstlisting} [basicstyle=\small, frame=single, breaklines=true, columns=fullflexible]

Name: AADSignInEventsBeta
Description: Information about Azure Active Directory (AAD) sign-in events either by a user (interactive) or a client on the user's behalf (non-interactive)
Columns: ['Timestamp', 'Application', 'ApplicationId', 'LogonType', 'EndpointCall', 'ErrorCode', 'CorrelationId', 'SessionId', 'AccountDisplayName', 'AccountObjectId', 'AccountUpn', 'IsExternalUser', 'IsGuestUser', 'AlternateSignInName', 'LastPasswordChangeTimestamp', 'ResourceDisplayName', 'ResourceId', 'ResourceTenantId', 'DeviceName', 'AadDeviceId', 'OSPlatform', 'DeviceTrustType', 'IsManaged', 'IsCompliant', 'AuthenticationProcessingDetails', 'AuthenticationRequirement', 'TokenIssuerType', 'RiskLevelAggregated', 'RiskLevelDuringSignIn', 'RiskEventTypes', 'RiskState', 'UserAgent', 'ClientAppUsed', 'Browser', 'ConditionalAccessPolicies', 'ConditionalAccessStatus', 'IPAddress', 'Country', 'State', 'City', 'Latitude', 'Longitude', 'NetworkLocationDetails', 'RequestId', 'ReportId', 'OAUTHProtocol']

Name: AADSpnSignInEventsBeta
Description: Information about sign-in events initiated by Azure Active Directory (AAD) service principal or managed identities
Columns: ['Timestamp', 'Application', 'ApplicationId', 'IsManagedIdentity', 'ErrorCode', 'CorrelationId', 'ServicePrincipalName', 'ServicePrincipalId', 'ResourceDisplayName', 'ResourceId', 'ResourceTenantId', 'IPAddress', 'Country', 'State', 'City', 'Latitude', 'Longitude', 'RequestId', 'ReportId']

Name: AlertEvidence
Description: Information about alerts from Microsoft Defender for Endpoint, Microsoft Defender for Office 365, Microsoft Cloud App Security, and Microsoft Defender for Identity, including severity information and threat categorization and the files, emails, IP addresses, URLs, users, devices or other entities associated with them. Each piece of evidence for an alert will have its own row with only the appropriate columns populated.
Columns: ['Timestamp', 'AlertId', 'Title', 'Categories', 'AttackTechniques', 'ServiceSource', 'DetectionSource', 'EntityType', 'EvidenceRole', 'EvidenceDirection', 'FileName', 'FolderPath', 'SHA1', 'SHA256', 'FileSize', 'ThreatFamily', 'RemoteIP', 'RemoteUrl', 'AccountName', 'AccountDomain', 'AccountSid', 'AccountObjectId', 'AccountUpn', 'DeviceId', 'DeviceName', 'LocalIP', 'NetworkMessageId', 'EmailSubject', 'Application', 'ApplicationId', 'OAuthApplicationId', 'ProcessCommandLine', 'RegistryKey', 'RegistryValueName', 'RegistryValueData', 'AdditionalFields', 'Severity', 'CloudPlatform', 'CloudResource', 'EmailClusterId', 'ResourceID', 'ResourceType', 'SubscriptionId']

Name: AlertInfo
Description: Alerts from Microsoft Defender for Endpoint, Microsoft Defender for Office 365, Microsoft Cloud App Security, and Microsoft Defender for Identity, including severity information and threat categorization
Columns: ['Timestamp', 'AlertId', 'Title', 'Category', 'Severity', 'ServiceSource', 'DetectionSource', 'AttackTechniques', 'AdditionalFields']

Name: CloudAppEvents
Description: Events involving accounts and objects in Office 365 and other cloud apps and services
Columns: ['Timestamp', 'ActionType', 'Application', 'ApplicationId', 'AppInstanceId', 'AccountObjectId', 'AccountId', 'AccountDisplayName', 'IsAdminOperation', 'DeviceType', 'OSPlatform', 'IPAddress', 'IsAnonymousProxy', 'CountryCode', 'City', 'ISP', 'UserAgent', 'ActivityType', 'ActivityObjects', 'ObjectName', 'ObjectType', 'ObjectId', 'ReportId', 'AccountType', 'IsExternalUser', 'IsImpersonated', 'IPTags', 'IPCategory', 'UserAgentTags', 'RawEventData', 'AdditionalFields', 'FirstSeen', 'LastSeenForUser', 'ThreatIndicators', 'UncommonForUser']

Name: DeviceEvents
Description: Multiple event types, including events triggered by security controls such as Windows Defender Antivirus and exploit protection
Columns: ['Timestamp', 'DeviceId', 'DeviceName', 'ActionType', 'FileName', 'FolderPath', 'SHA1', 'SHA256', 'MD5', 'FileSize', 'AccountDomain', 'AccountName', 'AccountSid', 'RemoteUrl', 'RemoteDeviceName', 'ProcessId', 'ProcessCommandLine', 'ProcessCreationTime', 'ProcessTokenElevation', 'LogonId', 'RegistryKey', 'RegistryValueName', 'RegistryValueData', 'RemoteIP', 'RemotePort', 'LocalIP', 'LocalPort', 'FileOriginUrl', 'FileOriginIP', 'InitiatingProcessSHA1', 'InitiatingProcessSHA256', 'InitiatingProcessMD5', 'InitiatingProcessFileName', 'InitiatingProcessFileSize', 'InitiatingProcessFolderPath', 'InitiatingProcessId', 'InitiatingProcessCommandLine', 'InitiatingProcessCreationTime', 'InitiatingProcessAccountDomain', 'InitiatingProcessAccountName', 'InitiatingProcessAccountSid', 'InitiatingProcessAccountUpn', 'InitiatingProcessAccountObjectId', 'InitiatingProcessVersionInfoCompanyName', 'InitiatingProcessVersionInfoProductName', 'InitiatingProcessVersionInfoProductVersion', 'InitiatingProcessVersionInfoInternalFileName', 'InitiatingProcessVersionInfoOriginalFileName', 'InitiatingProcessVersionInfoFileDescription', 'InitiatingProcessParentId', 'InitiatingProcessParentFileName', 'InitiatingProcessParentCreationTime', 'InitiatingProcessLogonId', 'ReportId', 'AppGuardContainerId', 'AdditionalFields']

Name: DeviceFileCertificateInfo
Description: Certificate information of signed files obtained from certificate verification events on endpoints
Columns: ['Timestamp', 'DeviceId', 'DeviceName', 'SHA1', 'IsSigned', 'SignatureType', 'Signer', 'SignerHash', 'Issuer', 'IssuerHash', 'CertificateSerialNumber', 'CrlDistributionPointUrls', 'CertificateCreationTime', 'CertificateExpirationTime', 'CertificateCountersignatureTime', 'IsTrusted', 'IsRootSignerMicrosoft', 'ReportId']

Name: DeviceFileEvents
Description: File creation, modification, and other file system events
Columns: ['Timestamp', 'DeviceId', 'DeviceName', 'ActionType', 'FileName', 'FolderPath', 'SHA1', 'SHA256', 'MD5', 'FileOriginUrl', 'FileOriginReferrerUrl', 'FileOriginIP', 'PreviousFolderPath', 'PreviousFileName', 'FileSize', 'InitiatingProcessAccountDomain', 'InitiatingProcessAccountName', 'InitiatingProcessAccountSid', 'InitiatingProcessAccountUpn', 'InitiatingProcessAccountObjectId', 'InitiatingProcessMD5', 'InitiatingProcessSHA1', 'InitiatingProcessSHA256', 'InitiatingProcessFolderPath', 'InitiatingProcessFileName', 'InitiatingProcessFileSize', 'InitiatingProcessVersionInfoCompanyName', 'InitiatingProcessVersionInfoProductName', 'InitiatingProcessVersionInfoProductVersion', 'InitiatingProcessVersionInfoInternalFileName', 'InitiatingProcessVersionInfoOriginalFileName', 'InitiatingProcessVersionInfoFileDescription', 'InitiatingProcessId', 'InitiatingProcessCommandLine', 'InitiatingProcessCreationTime', 'InitiatingProcessIntegrityLevel', 'InitiatingProcessTokenElevation', 'InitiatingProcessParentId', 'InitiatingProcessParentFileName', 'InitiatingProcessParentCreationTime', 'RequestProtocol', 'RequestSourceIP', 'RequestSourcePort', 'RequestAccountName', 'RequestAccountDomain', 'RequestAccountSid', 'ShareName', 'SensitivityLabel', 'SensitivitySubLabel', 'IsAzureInfoProtectionApplied', 'ReportId', 'AppGuardContainerId', 'AdditionalFields']

Name: DeviceImageLoadEvents
Description: DLL loading events on devices in the network, including information on the file loaded.
Columns: ['Timestamp', 'DeviceId', 'DeviceName', 'ActionType', 'FileName', 'FolderPath', 'SHA1', 'SHA256', 'MD5', 'FileSize', 'InitiatingProcessAccountDomain', 'InitiatingProcessAccountName', 'InitiatingProcessAccountSid', 'InitiatingProcessAccountUpn', 'InitiatingProcessAccountObjectId', 'InitiatingProcessIntegrityLevel', 'InitiatingProcessTokenElevation', 'InitiatingProcessSHA1', 'InitiatingProcessSHA256', 'InitiatingProcessMD5', 'InitiatingProcessFileName', 'InitiatingProcessFileSize', 'InitiatingProcessVersionInfoCompanyName', 'InitiatingProcessVersionInfoProductName', 'InitiatingProcessVersionInfoProductVersion', 'InitiatingProcessVersionInfoInternalFileName', 'InitiatingProcessVersionInfoOriginalFileName', 'InitiatingProcessVersionInfoFileDescription', 'InitiatingProcessId', 'InitiatingProcessCommandLine', 'InitiatingProcessCreationTime', 'InitiatingProcessFolderPath', 'InitiatingProcessParentId', 'InitiatingProcessParentFileName', 'InitiatingProcessParentCreationTime', 'ReportId', 'AppGuardContainerId']

Name: DeviceInfo
Description: Machine information, including OS information
Columns: ['Timestamp', 'DeviceId', 'DeviceName', 'ClientVersion', 'PublicIP', 'OSArchitecture', 'OSPlatform', 'OSBuild', 'IsAzureADJoined', 'JoinType', 'AadDeviceId', 'LoggedOnUsers', 'RegistryDeviceTag', 'OSVersion', 'MachineGroup', 'ReportId', 'OnboardingStatus', 'AdditionalFields', 'DeviceCategory', 'DeviceType', 'DeviceSubtype', 'Model', 'Vendor', 'OSDistribution', 'OSVersionInfo', 'MergedDeviceIds', 'MergedToDeviceId', 'AssetValue', 'DeviceDynamicTags', 'DeviceManualTags', 'DeviceRole', 'ExclusionReason', 'ExposureLevel', 'IsExcluded', 'IsInternetFacing', 'MitigationStatus', 'SensorHealthState']

Name: DeviceLogonEvents
Description: Sign-ins and other authentication events on machines, including the type of sign-in and the process that initiated the sign-in, where applicable.
Columns: ['Timestamp', 'DeviceId', 'DeviceName', 'ActionType', 'LogonType', 'AccountDomain', 'AccountName', 'AccountSid', 'Protocol', 'FailureReason', 'IsLocalAdmin', 'LogonId', 'RemoteDeviceName', 'RemoteIP', 'RemoteIPType', 'RemotePort', 'InitiatingProcessAccountDomain', 'InitiatingProcessAccountName', 'InitiatingProcessAccountSid', 'InitiatingProcessAccountUpn', 'InitiatingProcessAccountObjectId', 'InitiatingProcessIntegrityLevel', 'InitiatingProcessTokenElevation', 'InitiatingProcessSHA1', 'InitiatingProcessSHA256', 'InitiatingProcessMD5', 'InitiatingProcessFileName', 'InitiatingProcessFileSize', 'InitiatingProcessVersionInfoCompanyName', 'InitiatingProcessVersionInfoProductName', 'InitiatingProcessVersionInfoProductVersion', 'InitiatingProcessVersionInfoInternalFileName', 'InitiatingProcessVersionInfoOriginalFileName', 'InitiatingProcessVersionInfoFileDescription', 'InitiatingProcessId', 'InitiatingProcessCommandLine', 'InitiatingProcessCreationTime', 'InitiatingProcessFolderPath', 'InitiatingProcessParentId', 'InitiatingProcessParentFileName', 'InitiatingProcessParentCreationTime', 'ReportId', 'AppGuardContainerId', 'AdditionalFields']

Name: DeviceNetworkEvents
Description: Network connection and related events, with information on the local and remote IP addresses, type of connection, initiating process, URL or remote domain, and more.
Columns: ['Timestamp', 'DeviceId', 'DeviceName', 'ActionType', 'RemoteIP', 'RemotePort', 'RemoteUrl', 'LocalIP', 'LocalPort', 'Protocol', 'LocalIPType', 'RemoteIPType', 'InitiatingProcessSHA1', 'InitiatingProcessSHA256', 'InitiatingProcessMD5', 'InitiatingProcessFileName', 'InitiatingProcessFileSize', 'InitiatingProcessVersionInfoCompanyName', 'InitiatingProcessVersionInfoProductName', 'InitiatingProcessVersionInfoProductVersion', 'InitiatingProcessVersionInfoInternalFileName', 'InitiatingProcessVersionInfoOriginalFileName', 'InitiatingProcessVersionInfoFileDescription', 'InitiatingProcessId', 'InitiatingProcessCommandLine', 'InitiatingProcessCreationTime', 'InitiatingProcessFolderPath', 'InitiatingProcessParentFileName', 'InitiatingProcessParentId', 'InitiatingProcessParentCreationTime', 'InitiatingProcessAccountDomain', 'InitiatingProcessAccountName', 'InitiatingProcessAccountSid', 'InitiatingProcessAccountUpn', 'InitiatingProcessAccountObjectId', 'InitiatingProcessIntegrityLevel', 'InitiatingProcessTokenElevation', 'ReportId', 'AppGuardContainerId', 'AdditionalFields']

Name: DeviceNetworkInfo
Description: Network properties of machines, including adapters, IP and MAC addresses, as well as connected networks and domains
Columns: ['Timestamp', 'DeviceId', 'DeviceName', 'NetworkAdapterName', 'MacAddress', 'NetworkAdapterType', 'NetworkAdapterStatus', 'TunnelType', 'ConnectedNetworks', 'DnsAddresses', 'IPv4Dhcp', 'IPv6Dhcp', 'DefaultGateways', 'IPAddresses', 'ReportId', 'NetworkAdapterVendor']

Name: DeviceProcessEvents
Description: Process creation and related events including information on the associated users, devices, and files in addition to the process itself such as its command line, elevation, and more.
Columns: ['Timestamp', 'DeviceId', 'DeviceName', 'ActionType', 'FileName', 'FolderPath', 'SHA1', 'SHA256', 'MD5', 'FileSize', 'ProcessVersionInfoCompanyName', 'ProcessVersionInfoProductName', 'ProcessVersionInfoProductVersion', 'ProcessVersionInfoInternalFileName', 'ProcessVersionInfoOriginalFileName', 'ProcessVersionInfoFileDescription', 'ProcessId', 'ProcessCommandLine', 'ProcessIntegrityLevel', 'ProcessTokenElevation', 'ProcessCreationTime', 'AccountDomain', 'AccountName', 'AccountSid', 'AccountUpn', 'AccountObjectId', 'LogonId', 'InitiatingProcessAccountDomain', 'InitiatingProcessAccountName', 'InitiatingProcessAccountSid', 'InitiatingProcessAccountUpn', 'InitiatingProcessAccountObjectId', 'InitiatingProcessLogonId', 'InitiatingProcessIntegrityLevel', 'InitiatingProcessTokenElevation', 'InitiatingProcessSHA1', 'InitiatingProcessSHA256', 'InitiatingProcessMD5', 'InitiatingProcessFileName', 'InitiatingProcessFileSize', 'InitiatingProcessVersionInfoCompanyName', 'InitiatingProcessVersionInfoProductName', 'InitiatingProcessVersionInfoProductVersion', 'InitiatingProcessVersionInfoInternalFileName', 'InitiatingProcessVersionInfoOriginalFileName', 'InitiatingProcessVersionInfoFileDescription', 'InitiatingProcessId', 'InitiatingProcessCommandLine', 'InitiatingProcessCreationTime', 'InitiatingProcessFolderPath', 'InitiatingProcessParentId', 'InitiatingProcessParentFileName', 'InitiatingProcessParentCreationTime', 'InitiatingProcessSignerType', 'InitiatingProcessSignatureStatus', 'ReportId', 'AppGuardContainerId', 'AdditionalFields']

Name: DeviceRegistryEvents
Description: Creation, deletion, and modification of registry entries, including the actor and device involved, previous and current values, and more. This information can be used to track and investigate changes to a machine's configuration.
Columns: ['Timestamp', 'DeviceId', 'DeviceName', 'ActionType', 'RegistryKey', 'RegistryValueType', 'RegistryValueName', 'RegistryValueData', 'PreviousRegistryKey', 'PreviousRegistryValueName', 'PreviousRegistryValueData', 'InitiatingProcessAccountDomain', 'InitiatingProcessAccountName', 'InitiatingProcessAccountSid', 'InitiatingProcessAccountUpn', 'InitiatingProcessAccountObjectId', 'InitiatingProcessSHA1', 'InitiatingProcessSHA256', 'InitiatingProcessMD5', 'InitiatingProcessFileName', 'InitiatingProcessFileSize', 'InitiatingProcessVersionInfoCompanyName', 'InitiatingProcessVersionInfoProductName', 'InitiatingProcessVersionInfoProductVersion', 'InitiatingProcessVersionInfoInternalFileName', 'InitiatingProcessVersionInfoOriginalFileName', 'InitiatingProcessVersionInfoFileDescription', 'InitiatingProcessId', 'InitiatingProcessCommandLine', 'InitiatingProcessCreationTime', 'InitiatingProcessFolderPath', 'InitiatingProcessParentId', 'InitiatingProcessParentFileName', 'InitiatingProcessParentCreationTime', 'InitiatingProcessIntegrityLevel', 'InitiatingProcessTokenElevation', 'ReportId', 'AppGuardContainerId']

Name: DeviceTvmHardwareFirmware
Description: Information about device hardware and firmware, as checked by Microsoft Defender Vulnerability Management. The information includes the system model, processor, and BIOS, among others.
Columns: ['DeviceId', 'DeviceName', 'ComponentType', 'Manufacturer', 'ComponentName', 'ComponentFamily', 'ComponentVersion', 'AdditionalFields']

Name: DeviceTvmSecureConfigurationAssessment
Description: Threat & Vulnerability Management assessment events, indicating the status of various security configurations on devices
Columns: ['DeviceId', 'DeviceName', 'OSPlatform', 'Timestamp', 'ConfigurationId', 'ConfigurationCategory', 'ConfigurationSubcategory', 'ConfigurationImpact', 'IsCompliant', 'IsApplicable', 'Context', 'IsExpectedUserImpact']

Name: DeviceTvmSoftwareInventory
Description: Inventory of software installed on devices, including their version information and end-of-support status
Columns: ['DeviceId', 'DeviceName', 'OSPlatform', 'OSVersion', 'OSArchitecture', 'SoftwareVendor', 'SoftwareName', 'SoftwareVersion', 'EndOfSupportStatus', 'EndOfSupportDate', 'ProductCodeCpe']

Name: DeviceTvmSoftwareVulnerabilities
Description: Software vulnerabilities found on devices and the list of available security updates that address each vulnerability
Columns: ['DeviceId', 'DeviceName', 'OSPlatform', 'OSVersion', 'OSArchitecture', 'SoftwareVendor', 'SoftwareName', 'SoftwareVersion', 'CveId', 'VulnerabilitySeverityLevel', 'RecommendedSecurityUpdate', 'RecommendedSecurityUpdateId', 'CveTags', 'CveMitigationStatus']

Name: DeviceTvmSoftwareVulnerabilitiesKB
Description: Knowledge base of publicly disclosed vulnerabilities, including whether exploit code is publicly available
Columns: ['CveId', 'CvssScore', 'IsExploitAvailable', 'VulnerabilitySeverityLevel', 'LastModifiedTime', 'PublishedDate', 'VulnerabilityDescription', 'AffectedSoftware']

Name: EmailAttachmentInfo
Description: Information about files attached to Office 365 emails, including details about the containing email and any threats detected in the attachment
Columns: ['Timestamp', 'NetworkMessageId', 'SenderFromAddress', 'SenderDisplayName', 'SenderObjectId', 'RecipientEmailAddress', 'RecipientObjectId', 'FileName', 'FileType', 'SHA256', 'FileSize', 'ThreatTypes', 'ThreatNames', 'DetectionMethods', 'ReportId']

Name: EmailEvents
Description: Office 365 email events, including email delivery and blocking events, sender and recipient details, threat detection, and more.
Columns: ['Timestamp', 'NetworkMessageId', 'InternetMessageId', 'SenderMailFromAddress', 'SenderFromAddress', 'SenderDisplayName', 'SenderObjectId', 'SenderMailFromDomain', 'SenderFromDomain', 'SenderIPv4', 'SenderIPv6', 'RecipientEmailAddress', 'RecipientObjectId', 'Subject', 'EmailClusterId', 'EmailDirection', 'DeliveryAction', 'DeliveryLocation', 'ThreatTypes', 'ThreatNames', 'DetectionMethods', 'ConfidenceLevel', 'BulkComplaintLevel', 'EmailAction', 'EmailActionPolicy', 'EmailActionPolicyGuid', 'AuthenticationDetails', 'AttachmentCount', 'UrlCount', 'EmailLanguage', 'Connectors', 'OrgLevelAction', 'OrgLevelPolicy', 'UserLevelAction', 'UserLevelPolicy', 'ReportId', 'AdditionalFields', 'Cc', 'DistributionList', 'ExchangeTransportRule', 'ForwardingInformation', 'LatestDeliveryAction', 'LatestDeliveryLocation', 'To']

Name: EmailPostDeliveryEvents
Description: Security events that occur post-delivery, after Office 365 has delivered an email message to the recipient mailbox
Columns: ['Timestamp', 'NetworkMessageId', 'InternetMessageId', 'Action', 'ActionType', 'ActionTrigger', 'ActionResult', 'RecipientEmailAddress', 'DeliveryLocation', 'ThreatTypes', 'DetectionMethods', 'ReportId']

Name: EmailUrlInfo
Description: Information about URLs found within Office 365 emails
Columns: ['Timestamp', 'NetworkMessageId', 'Url', 'UrlDomain', 'UrlLocation', 'ReportId']

Name: IdentityDirectoryEvents
Description: Events involving an on-premises domain controller running Active Directory (AD). This table captures various identity-related events, like password changes, password expiration, and user principal name (UPN) changes. It also captures system events on the domain controller, like scheduling of tasks and PowerShell activity.
Columns: ['Timestamp', 'ActionType', 'Application', 'TargetAccountUpn', 'TargetAccountDisplayName', 'TargetDeviceName', 'DestinationDeviceName', 'DestinationIPAddress', 'DestinationPort', 'Protocol', 'AccountName', 'AccountDomain', 'AccountUpn', 'AccountSid', 'AccountObjectId', 'AccountDisplayName', 'DeviceName', 'IPAddress', 'Port', 'Location', 'ISP', 'ReportId', 'AdditionalFields']

Name: IdentityInfo
Description: Account information from various sources, including Azure Active Directory
Columns: ['Timestamp', 'AccountObjectId', 'AccountUpn', 'OnPremSid', 'AccountDisplayName', 'AccountName', 'AccountDomain', 'DistinguishedName', 'CloudSid', 'GivenName', 'Surname', 'Department', 'JobTitle', 'EmailAddress', 'SipProxyAddress', 'Address', 'City', 'Country', 'IsAccountEnabled', 'Manager', 'Phone', 'CreatedDateTime']

Name: IdentityLogonEvents
Description: Authentication activities made through an on-premises Active Directory captured by Microsoft Defender for Identity and other authentication activities related to Microsoft online services captured by Microsoft Defender for Cloud Apps.
Columns: ['Timestamp', 'ActionType', 'Application', 'LogonType', 'Protocol', 'FailureReason', 'AccountName', 'AccountDomain', 'AccountUpn', 'AccountSid', 'AccountObjectId', 'AccountDisplayName', 'DeviceName', 'DeviceType', 'OSPlatform', 'IPAddress', 'Port', 'DestinationDeviceName', 'DestinationIPAddress', 'DestinationPort', 'TargetDeviceName', 'TargetAccountDisplayName', 'Location', 'ISP', 'ReportId', 'AdditionalFields', 'FirstSeen', 'IPCategory', 'IPTags', 'LastSeenForUser', 'RawEventData', 'ThreatIndicators', 'UncommonForUser', 'UserAgent']

Name: IdentityQueryEvents
Description: Query activities performed against Active Directory objects, such as users, groups, devices, and domains
Columns: ['Timestamp', 'ActionType', 'Application', 'QueryType', 'QueryTarget', 'Query', 'Protocol', 'AccountName', 'AccountDomain', 'AccountUpn', 'AccountSid', 'AccountObjectId', 'AccountDisplayName', 'DeviceName', 'IPAddress', 'Port', 'DestinationDeviceName', 'DestinationIPAddress', 'DestinationPort', 'TargetDeviceName', 'TargetAccountUpn', 'TargetAccountDisplayName', 'Location', 'ReportId', 'AdditionalFields']

Name: UrlClickEvents
Description: Events involving URLs clicked, selected, or requested on Microsoft Defender for Office 365 from email messages, Microsoft Teams, and Office 365 apps in supported desktop, mobile, and web apps.
Columns: ['Timestamp', 'Url', 'ActionType', 'AccountUpn', 'Workload', 'NetworkMessageId', 'ThreatTypes', 'DetectionMethods', 'IPAddress', 'IsClickedThrough', 'UrlChain', 'ReportId']


\end{lstlisting}

\subsection{Sentinel Database}

\begin{lstlisting} [basicstyle=\small, frame=single, breaklines=true, columns=fullflexible]

Name: AADManagedIdentitySignInLogs
Description: Managed identity Azure Active Directory sign-in logs, including details such as app ID, authentication context, IP address, location, resource information, service principal details, and the result of the sign-in operation.
Columns: ['AppId', 'AuthenticationContextClassReferences', 'AuthenticationProcessingDetails', '_BilledSize', 'Category', 'ConditionalAccessPolicies', 'ConditionalAccessStatus', 'CorrelationId', 'DurationMs', 'FederatedCredentialId', 'Id', 'Identity', 'IPAddress', '_IsBillable', 'Level', 'Location', 'LocationDetails', 'OperationName', 'OperationVersion', 'ResourceDisplayName', 'ResourceGroup', 'ResourceIdentity', 'ResourceServicePrincipalId', 'ResultDescription', 'ResultSignature', 'ResultType', 'ServicePrincipalCredentialKeyId', 'ServicePrincipalCredentialThumbprint', 'ServicePrincipalId', 'ServicePrincipalName', 'SourceSystem', 'TenantId', 'TimeGenerated', 'Type', 'UniqueTokenIdentifier']

Name: AADNonInteractiveUserSignInLogs
Description: Non-interactive Azure Active Directory sign-in logs from users, including information on user identities, authentication methods, device details, IP addresses, risk levels, and sign-in statuses. It also provides details on conditional access policies, multi-factor authentication, and location information for each sign-in event.
Columns: ['AlternateSignInName', 'AppDisplayName', 'AppId', 'AppliedEventListeners', 'AuthenticationContextClassReferences', 'AuthenticationDetails', 'AuthenticationMethodsUsed', 'AuthenticationProcessingDetails', 'AuthenticationProtocol', 'AuthenticationRequirement', 'AuthenticationRequirementPolicies', 'AutonomousSystemNumber', '_BilledSize', 'Category', 'ClientAppUsed', 'ConditionalAccessPolicies', 'ConditionalAccessStatus', 'CorrelationId', 'CreatedDateTime', 'CrossTenantAccessType', 'DeviceDetail', 'DurationMs', 'HomeTenantId', 'Id', 'Identity', 'IPAddress', '_IsBillable', 'IsInteractive', 'IsRisky', 'Level', 'Location', 'LocationDetails', 'MfaDetail', 'NetworkLocationDetails', 'OperationName', 'OperationVersion', 'OriginalRequestId', 'ProcessingTimeInMs', 'ResourceDisplayName', 'ResourceGroup', 'ResourceIdentity', 'ResourceServicePrincipalId', 'ResourceTenantId', 'ResultDescription', 'ResultSignature', 'ResultType', 'RiskDetail', 'RiskEventTypes', 'RiskEventTypes_V2', 'RiskLevelAggregated', 'RiskLevelDuringSignIn', 'RiskState', 'ServicePrincipalId', 'SessionLifetimePolicies', 'SignInEventTypes', 'SignInIdentifierType', 'SourceSystem', 'Status', 'TenantId', 'TimeGenerated', 'TokenIssuerName', 'TokenIssuerType', 'Type', 'UniqueTokenIdentifier', 'UserAgent', 'UserDisplayName', 'UserId', 'UserPrincipalName', 'UserType']

Name: AADRiskyUsers
Description: Details about Azure AD Risky Users identified by Identity Protection, including user IDs, risk details, risk levels, risk states, and the date and time of the event. It also provides information on whether the user is deleted, whether their risky state is being processed, and the user's display and principal names.
Columns: ['_BilledSize', 'CorrelationId', 'Id', '_IsBillable', 'IsDeleted', 'IsProcessing', 'OperationName', 'RiskDetail', 'RiskLastUpdatedDateTime', 'RiskLevel', 'SourceSystem', 'TenantId', 'TimeGenerated', 'Type', 'UserDisplayName', 'UserPrincipalName']

Name: AADServicePrincipalSignInLogs
Description: Sign-in logs for service principals in Azure Active Directory, including details such as app ID, authentication context, IP address, location, resource information, service principal credentials, and the result of the sign-in operation.
Columns: ['AppId', 'AuthenticationContextClassReferences', 'AuthenticationProcessingDetails', '_BilledSize', 'Category', 'ConditionalAccessPolicies', 'ConditionalAccessStatus', 'CorrelationId', 'DurationMs', 'FederatedCredentialId', 'Id', 'Identity', 'IPAddress', '_IsBillable', 'Level', 'Location', 'LocationDetails', 'OperationName', 'OperationVersion', 'ResourceDisplayName', 'ResourceGroup', 'ResourceIdentity', 'ResourceServicePrincipalId', 'ResultDescription', 'ResultSignature', 'ResultType', 'ServicePrincipalCredentialKeyId', 'ServicePrincipalCredentialThumbprint', 'ServicePrincipalId', 'ServicePrincipalName', 'SourceSystem', 'TenantId', 'TimeGenerated', 'Type', 'UniqueTokenIdentifier']

Name: AADUserRiskEvents
Description: User risk events in Azure Active Directory, containing the activity type, date and time of the risky activity, additional information in JSON format, correlation ID, detected risk details, risk level, risk state, source, and user information such as display name, user ID, and user principal name.
Columns: ['Activity', 'ActivityDateTime', 'AdditionalInfo', '_BilledSize', 'CorrelationId', 'DetectedDateTime', 'DetectionTimingType', 'Id', 'IpAddress', '_IsBillable', 'LastUpdatedDateTime', 'Location', 'OperationName', 'RequestId', 'RiskDetail', 'RiskEventType', 'RiskLevel', 'RiskState', 'Source', 'SourceSystem', 'TenantId', 'TimeGenerated', 'TokenIssuerType', 'Type', 'UserDisplayName', 'UserId', 'UserPrincipalName']

Name: Anomalies
Description: Data on anomalies generated by analytics rules, including details about the rule, algorithm, and explanations for the anomaly, as well as information on the devices, users, and locations involved. It also provides timestamps, scores, and insights about the activities, entities, and tactics related to the anomaly.
Columns: ['ActivityInsights', 'AnomalyDetails', 'AnomalyReasons', 'AnomalyTemplateId', 'AnomalyTemplateName', 'AnomalyTemplateVersion', '_BilledSize', 'Description', 'DestinationDevice', 'DestinationIpAddress', 'DestinationLocation', 'DeviceInsights', 'EndTime', 'Entities', 'ExtendedLinks', 'ExtendedProperties', 'Id', '_IsBillable', 'RuleConfigVersion', 'RuleId', 'RuleName', 'RuleStatus', 'Score', 'SourceDevice', 'SourceIpAddress', 'SourceLocation', 'SourceSystem', 'StartTime', 'Tactics', 'Techniques', 'TenantId', 'TimeGenerated', 'Type', 'UserInsights', 'UserName', 'UserPrincipalName', 'VendorName', 'WorkspaceId']

Name: AWSCloudTrail
Description: Data and management events of an Amazon Web Services account, including information about API versions, event types, request and response details, error codes, user identities, and various other parameters related to AWS service actions and resources accessed.
Columns: ['AdditionalEventData', 'APIVersion', 'AwsEventId', 'AWSRegion', 'AwsRequestId_', 'AwsRequestId', '_BilledSize', 'Category', 'CipherSuite', 'ClientProvidedHostHeader', 'EC2RoleDelivery', 'ErrorCode', 'ErrorMessage', 'EventName', 'EventSource', 'EventTypeName', 'EventVersion', '_IsBillable', 'ManagementEvent', 'OperationName', 'ReadOnly', 'RecipientAccountId', 'RequestParameters', 'Resources', 'ResponseElements', 'ServiceEventDetails', 'SessionCreationDate', 'SessionIssuerAccountId', 'SessionIssuerArn', 'SessionIssuerPrincipalId', 'SessionIssuerType', 'SessionIssuerUserName', 'SessionMfaAuthenticated', 'SharedEventId', 'SourceIpAddress', 'SourceSystem', 'TenantId', 'TimeGenerated', 'TlsVersion', 'Type', 'UserAgent', 'UserIdentityAccessKeyId', 'UserIdentityAccountId', 'UserIdentityArn', 'UserIdentityInvokedBy', 'UserIdentityPrincipalid', 'UserIdentityType', 'UserIdentityUserName', 'VpcEndpointId']

Name: AWSGuardDuty
Description: Amazon GuardDuty findings for potential security issues detected within a network, including details on the AWS account, activity type, resource targeted, severity level, and timestamps for when the finding was created and last updated.
Columns: ['AccountId', 'ActivityType', 'Arn', '_BilledSize', 'Description', 'Id', '_IsBillable', 'Partition', 'Region', 'ResourceDetails', 'SchemaVersion', 'ServiceDetails', 'Severity', 'SourceSystem', 'TenantId', 'TimeCreated', 'TimeGenerated', 'Title', 'Type']

Name: BehaviorAnalytics
Description: Enriched events for behavior analytics, including information on action types, activity insights, actor names, devices, IP addresses, and locations. Additionally, it provides data on event sources, vendors, investigation priority scores, and user metadata insights.
Columns: ['ActionType', 'ActivityInsights', 'ActivityType', 'ActorName', 'ActorPrincipalName', '_BilledSize', 'DestinationDevice', 'DestinationIPAddress', 'DestinationIPLocation', 'Device', 'DevicesInsights', 'EventProductVersion', 'EventSource', 'EventVendor', 'InvestigationPriority', '_IsBillable', 'NativeTableName', '_ResourceId', 'SourceDevice', 'SourceIPAddress', 'SourceIPLocation', 'SourceRecordId', 'SourceSystem', '_SubscriptionId', 'TargetName', 'TargetPrincipalName', 'TenantId', 'TimeGenerated', 'TimeProcessed', 'Type', 'UserName', 'UserPrincipalName', 'UsersInsights']

Name: Event
Description: Data from Windows Event Log on Windows computers, including information such as event category, severity, message, source, and user name, as well as details about the computer, cloud service, and resource associated with the event. Additionally, it provides information on billing and management groups for the collected data.
Columns: ['AzureDeploymentID', '_BilledSize', 'Computer', 'EventCategory', 'EventData', 'EventID', 'EventLevel', 'EventLevelName', 'EventLog', '_IsBillable', 'ManagementGroupName', 'Message', 'ParameterXml', 'RenderedDescription', '_ResourceId', 'Role', 'Source', 'SourceSystem', '_SubscriptionId', 'TimeGenerated', 'Type', 'UserName']

Name: Heartbeat
Description: Data on Log Analytics agents' health, including information on the computer, operating system, agent version, and geographic location. Additionally, it provides details on the Azure resource running the agent, such as resource group, provider, and subscription ID.
Columns: ['_BilledSize', 'Category', 'Computer', 'ComputerEnvironment', 'ComputerIP', 'ComputerPrivateIPs', '_IsBillable', 'IsGatewayInstalled', 'ManagementGroupName', 'OSMajorVersion', 'OSMinorVersion', 'OSName', 'OSType', 'RemoteIPCountry', 'RemoteIPLatitude', 'RemoteIPLongitude', 'Resource', 'ResourceGroup', '_ResourceId', 'ResourceId', 'ResourceProvider', 'ResourceType', 'SCAgentChannel', 'Solutions', 'SourceSystem', 'SubscriptionId', '_SubscriptionId', 'TimeGenerated', 'Type', 'Version', 'VMUUID']

Name: IdentityInfo
Description: Covers user accounts with details like display names, email addresses, and job titles, as well as domain, location, and more. Offers insights into user profiles and can be used for user management or analyzing user behavior in a system.
Columns: ['AccountCloudSID', 'AccountCreationTime', 'AccountDisplayName', 'AccountDomain', 'AccountName', 'AccountObjectId', 'AccountSID', 'AccountTenantId', 'AccountUPN', 'AdditionalMailAddresses', 'Applications', 'AssignedRoles', '_BilledSize', 'BlastRadius', 'ChangeSource', 'City', 'CompanyName', 'Country', 'DeletedDateTime', 'Department', 'EmployeeId', 'EntityRiskScore', 'ExtensionProperty', 'GivenName', 'GroupMembership', 'InvestigationPriority', 'InvestigationPriorityPercentile', 'IsAccountEnabled', '_IsBillable', 'IsMFARegistered', 'IsServiceAccount', 'JobTitle', 'LastSeenDate', 'MailAddress', 'Manager', 'OnPremisesDistinguishedName', 'OnPremisesExtensionAttributes', 'Phone', 'RelatedAccounts', 'RiskLevel', 'RiskLevelDetails', 'RiskState', 'ServicePrincipals', 'SourceSystem', 'State', 'StreetAddress', 'Surname', 'Tags', 'TenantId', 'TimeGenerated', 'Type', 'UACFlags', 'UserAccountControl', 'UserState', 'UserStateChangedOn', 'UserType']

Name: SecurityIncident
Description: Information about security incidents, including details such as incident IDs, classifications, timestamps, descriptions, severity, status, and related alerts, bookmarks, and analytic rules. It also provides data on incident modifications, assigned users, source systems, and billing information.
Columns: ['AdditionalData', 'AlertIds', '_BilledSize', 'BookmarkIds', 'Classification', 'ClassificationComment', 'ClassificationReason', 'ClosedTime', 'Comments', 'CreatedTime', 'Description', 'FirstActivityTime', 'FirstModifiedTime', 'IncidentName', 'IncidentNumber', 'IncidentUrl', '_IsBillable', 'Labels', 'LastActivityTime', 'LastModifiedTime', 'ModifiedBy', 'Owner', 'ProviderIncidentId', 'ProviderName', 'RelatedAnalyticRuleIds', 'Severity', 'SourceSystem', 'Status', 'Tasks', 'TenantId', 'TimeGenerated', 'Title', 'Type']

Name: SqlAtpStatus
Description: Status information about SQL Advanced Threat Protection, including agent and instance details, protection status, errors, and billing information for machines connected to the workspace. It helps identify and diagnose issues with SQL ATP protection on each instance.
Columns: ['AgentId', 'AgentStartTime', '_BilledSize', 'ClientIP', 'Computer', 'HostResourceId', 'IntelligencePackVersion', '_IsBillable', 'LastError', 'MachineUUID', 'SourceSystem', 'SqlInstanceName', 'SqlInstanceStartTime', 'SqlInstanceVersion', 'Status', 'TenantId', 'TimeGenerated', 'Type']

Name: StorageBlobLogs
Description: Storage blob details and events, including storage account, authentication, authorization, operation type, request and response sizes, latency, and status codes. It also provides data on the requester's IP address, user agent, and OAuth details.
Columns: ['AccountName', 'AuthenticationHash', 'AuthenticationType', 'AuthorizationDetails', '_BilledSize', 'CallerIpAddress', 'Category', 'ClientRequestId', 'ConditionsUsed', 'ContentLengthHeader', 'CorrelationId', 'DurationMs', 'Etag', '_IsBillable', 'LastModifiedTime', 'Location', 'OperationCount', 'OperationName', 'OperationVersion', 'Protocol', 'ReferrerHeader', 'RequestBodySize', 'RequesterAppId', 'RequesterAudience', 'RequesterObjectId', 'RequesterTenantId', 'RequesterTokenIssuer', 'RequesterUpn', 'RequestHeaderSize', 'RequestMd5', '_ResourceId', 'ResponseBodySize', 'ResponseHeaderSize', 'ResponseMd5', 'SchemaVersion', 'ServerLatencyMs', 'ServiceType', 'SourceSystem', 'StatusCode', 'StatusText', '_SubscriptionId', 'TenantId', 'TimeGenerated', 'TlsVersion', 'Type', 'Uri', 'UserAgentHeader']

Name: StorageFileLogs
Description: Logs for storage file services, including details on the storage account, authentication, authorization, operation type, request and response sizes, protocol, and status codes. It also provides data on the requester's IP address, user agent, and OAuth details, as well as timestamps and service-related identifiers.
Columns: ['AccountName', 'AuthenticationHash', 'AuthenticationType', 'AuthorizationDetails', '_BilledSize', 'CallerIpAddress', 'Category', 'ClientRequestId', 'ConditionsUsed', 'ContentLengthHeader', 'CorrelationId', 'DurationMs', 'Etag', '_IsBillable', 'LastModifiedTime', 'Location', 'OperationCount', 'OperationName', 'OperationVersion', 'Protocol', 'ReferrerHeader', 'RequestBodySize', 'RequesterAppId', 'RequesterAudience', 'RequesterObjectId', 'RequesterTenantId', 'RequesterTokenIssuer', 'RequesterUpn', 'RequesterUserName', 'RequestHeaderSize', 'RequestMd5', '_ResourceId', 'ResponseBodySize', 'ResponseHeaderSize', 'ResponseMd5', 'SchemaVersion', 'ServerLatencyMs', 'ServiceType', 'SmbCommandDetail', 'SmbCommandMajor', 'SmbCommandMinor', 'SmbCreditsConsumed', 'SmbFileId', 'SmbMessageID', 'SmbPersistentHandleID', 'SmbPrimarySID', 'SmbSessionID', 'SmbTreeConnectID', 'SmbVolatileHandleID', 'SourceSystem', 'StatusCode', 'StatusText', '_SubscriptionId', 'TenantId', 'TimeGenerated', 'TlsVersion', 'Type', 'Uri', 'UserAgentHeader']

Name: Syslog
Description: Syslog events from Linux computers, including information such as the record size, computer name, event time, facility, host IP and name, process ID and name, severity level, and syslog message text. Additionally, it provides details about the source system, resource and subscription identifiers, and billing information.
Columns: ['_BilledSize', 'CollectorHostName', 'Computer', 'EventTime', 'Facility', 'HostIP', 'HostName', '_IsBillable', 'ProcessID', 'ProcessName', '_ResourceId', 'SeverityLevel', 'SourceSystem', '_SubscriptionId', 'SyslogMessage', 'TimeGenerated', 'Type']

Name: ThreatIntelligenceIndicator
Description: Information on threat intelligence indicators, including details on actions, activity groups, confidence scores, descriptions, domain names, email observables, file observables, network observables, malware names, threat severity, threat types, and traffic light protocol levels. It also provides information on whether the indicator is active, passive, or associated with known false positives.
Columns: ['Action', 'Active', 'ActivityGroupNames', 'AdditionalInformation', '_BilledSize', 'ConfidenceScore', 'Description', 'DiamondModel', 'DomainName', 'EmailEncoding', 'EmailLanguage', 'EmailRecipient', 'EmailSenderAddress', 'EmailSenderName', 'EmailSourceDomain', 'EmailSourceIpAddress', 'EmailSubject', 'EmailXMailer', 'ExpirationDateTime', 'ExternalIndicatorId', 'FileCompileDateTime', 'FileCreatedDateTime', 'FileHashType', 'FileHashValue', 'FileMutexName', 'FileName', 'FilePacker', 'FilePath', 'FileSize', 'FileType', 'IndicatorId', 'IndicatorProvider', '_IsBillable', 'KillChainActions', 'KillChainC2', 'KillChainDelivery', 'KillChainExploitation', 'KillChainReconnaissance', 'KillChainWeaponization', 'KnownFalsePositives', 'MalwareNames', 'NetworkCidrBlock', 'NetworkDestinationAsn', 'NetworkDestinationCidrBlock', 'NetworkDestinationIP', 'NetworkDestinationPort', 'NetworkIP', 'NetworkPort', 'NetworkProtocol', 'NetworkSourceAsn', 'NetworkSourceCidrBlock', 'NetworkSourceIP', 'NetworkSourcePort', 'PassiveOnly', 'Tags', 'TenantId', 'ThreatSeverity', 'ThreatType', 'TimeGenerated', 'TrafficLightProtocolLevel', 'Type', 'Url', 'UserAgent']

Name: VMComputer
Description: Inventory data for servers, including information on agent IDs, Azure services, VM details, boot time, CPU specifications, DNS names, IP addresses, MAC addresses, operating system details, physical memory, and virtualization state.
Columns: ['AgentId', 'AzureCloudServiceDeployment', 'AzureCloudServiceInstanceId', 'AzureCloudServiceName', 'AzureCloudServiceRoleName', 'AzureCloudServiceRoleType', 'AzureFaultDomain', 'AzureImageOffering', 'AzureImagePublisher', 'AzureImageSku', 'AzureImageVersion', 'AzureLocation', 'AzureResourceGroup', 'AzureResourceName', 'AzureServiceFabricClusterId', 'AzureServiceFabricClusterName', 'AzureSize', 'AzureSubscriptionId', 'AzureUpdateDomain', 'AzureVmId', 'AzureVmScaleSetDeployment', 'AzureVmScaleSetInstanceId', 'AzureVmScaleSetName', 'AzureVmScaleSetResourceId', '_BilledSize', 'BootTime', 'Computer', 'Cpus', 'CpuSpeed', 'DependencyAgentVersion', 'DisplayName', 'DnsNames', 'FullDisplayName', 'HostingProvider', 'HostName', 'HypervisorId', 'HypervisorType', 'Ipv4Addresses', 'Ipv4DefaultGateways', 'Ipv4SubnetMasks', 'Ipv6Addresses', '_IsBillable', 'MacAddresses', 'Machine', 'OperatingSystemFamily', 'OperatingSystemFullName', 'PhysicalMemoryMB', '_ResourceId', 'SourceSystem', '_SubscriptionId', 'TimeGenerated', 'TimeZone', 'Type', 'VirtualizationState', 'VirtualMachineHypervisorId', 'VirtualMachineNativeId', 'VirtualMachineNativeName', 'VirtualMachineType']

Name: VMBoundPort
Description: Data on open server ports, including information on bytes sent and received, port IP addresses, links established, terminated, and live, as well as response times and associated processes. It also provides unique identifiers for agents, machines, and processes, along with subscription and resource details.
Columns: ['AgentId', '_BilledSize', 'BytesReceived', 'BytesSent', 'Computer', 'Ip', '_IsBillable', 'IsWildcardBind', 'LinksEstablished', 'LinksLive', 'LinksTerminated', 'Machine', 'Port', 'PortId', 'Process', 'ProcessName', 'Protocol', '_ResourceId', 'Responses', 'ResponseTimeMax', 'ResponseTimeMin', 'ResponseTimeSum', 'SourceSystem', '_SubscriptionId', 'TimeGenerated', 'Type']

Name: VMConnection
Description: Data on inbound and outbound connections to and from monitored computers, including information on bytes sent and received, connection direction, IP addresses, ports, protocols, response times, and geolocation. Additionally, it provides details on observed threats, their severity, and Traffic Light Protocol (TLP) levels.
Columns: ['AgentId', '_BilledSize', 'BytesReceived', 'BytesSent', 'Computer', 'Confidence', 'ConnectionId', 'Description', 'DestinationIp', 'DestinationPort', 'Direction', 'FirstReportedDateTime', 'IndicatorThreatType', 'IsActive', '_IsBillable', 'LastReportedDateTime', 'LinksEstablished', 'LinksFailed', 'LinksLive', 'LinksTerminated', 'Machine', 'MaliciousIp', 'Process', 'ProcessName', 'Protocol', 'RemoteClassification', 'RemoteCountry', 'RemoteDnsCanonicalNames', 'RemoteDnsQuestions', 'RemoteIp', 'RemoteLatitude', 'RemoteLongitude', '_ResourceId', 'Responses', 'ResponseTimeMax', 'ResponseTimeMin', 'ResponseTimeSum', 'Severity', 'SourceIp', 'SourceSystem', '_SubscriptionId', 'TimeGenerated', 'TLPLevel', 'Type']

Name: VMProcess
Description: Process data for servers, including information about the agent, command line, company, computer, process description, executable details, file and product versions, process group, machine, services, start time, and user details.
Columns: ['AgentId', '_BilledSize', 'CommandLine', 'CompanyName', 'Computer', 'Description', 'DisplayName', 'ExecutableName', 'ExecutablePath', 'FileVersion', 'FirstPid', 'Group', 'InternalName', '_IsBillable', 'Machine', 'Process', 'ProductName', 'ProductVersion', '_ResourceId', 'Role', 'Services', 'SourceSystem', 'StartTime', '_SubscriptionId', 'TimeGenerated', 'Type', 'UserDomain', 'UserName', 'WorkingDirectory']

Name: W3CIISLog
Description: Data related to Internet Information Server (IIS) logs on Windows computers, including information about client IP addresses, server IP addresses, request methods, user agents, and more. It also provides details about the size of data sent and received, as well as location data for the client IP address.
Columns: ['AzureDeploymentID', '_BilledSize', 'cIP', 'Computer', 'Confidence', 'csBytes', 'csCookie', 'csHost', 'csMethod', 'csReferer', 'csUriQuery', 'csUriStem', 'csUserAgent', 'csUserName', 'csVersion', 'Description', 'FirstReportedDateTime', 'IndicatorThreatType', 'IsActive', '_IsBillable', 'LastReportedDateTime', 'MaliciousIP', 'ManagementGroupName', 'RemoteIPCountry', 'RemoteIPLatitude', 'RemoteIPLongitude', '_ResourceId', 'Role', 'RoleInstance', 'scBytes', 'scStatus', 'scSubStatus', 'scWin32Status', 'Severity', 'sIP', 'SourceSystem', 'sPort', 'sSiteName', 'StorageAccount', '_SubscriptionId', 'TimeGenerated', 'TimeTaken', 'TLPLevel', 'Type']

\end{lstlisting}

\end{document}